\documentclass[11pt,a4paper]{article}
\usepackage{jheppub}
\usepackage{amsmath,amssymb}
\usepackage{ulem}

\usepackage[splitrule]{footmisc} 
\global\interfootnotelinepenalty=10000

\graphicspath{{graphics/}}

\title{
Hydrodynamics of simply spinning black holes \& 
hydrodynamics for spinning quantum fluids
}
\author[]{Markus Garbiso and Matthias Kaminski}
\affiliation[]{Department of Physics and Astronomy, University of Alabama,\\ Tuscaloosa, AL 35487, USA}
\emailAdd{magarbiso@crimson.ua.edu}
\emailAdd{mski@ua.edu}

\begin{document}
\newcommand{\eref}[1]{Eq.~(\ref{#1})}
\newcommand{\exd}{\mathrm{d}}
\newcommand{\reals}{\mathbb{R}}

\abstract{ 
We find hydrodynamic behavior in large simply spinning five-dimensional Anti-de Sitter black holes. These are dual to spinning quantum fluids through the AdS/CFT correspondence constructed from string theory. 
Due to the spatial anisotropy introduced by the angular momentum, hydrodynamic transport coefficients are split into groups longitudinal or transverse to the angular momentum, and aligned or anti-aligned with it. 
Analytic expressions are provided for the two shear viscosities,
the longitudinal momentum diffusion coefficient,
two speeds of sound,
and two sound attenuation coefficients. 
Known relations between these coefficients are generalized to include dependence on angular momentum. 
The shear viscosity to entropy density ratio varies between zero and $1/(4\pi)$ depending on the direction of
the shear.
These results can be applied to heavy ion collisions, in which the most vortical fluid was reported recently. 
In passing, we show that large simply spinning five-dimensional Myers-Perry black holes are perturbatively stable for all angular momenta below extremality.  
}

\maketitle
\section{Introduction}
\label{sec:introduction}
%

In this work, we consider spinning black holes in five-dimensional spacetime which is asymptotic to Anti-de Sitter space~(AdS). We describe in which way this gravitational setup is holographically dual~\cite{Maldacena:1997re} to a spinning quantum fluid with nonzero vorticity and angular momentum. On one hand, this reveals new properties of simply spinning\footnote{In five dimensions there are two angular momenta characterizing the rotation of a black hole. With {\it simply spinning} we refer to the case in which these two angular momenta are equal to each other. This simplifies the analysis and retains the relevant effects to be studied here.} 
Myers-Perry black holes in $AdS_5$~\cite{Myers:1986un}. On the other hand, lessons are learned for the general hydrodynamic description of spinning quantum fluids. 

Such a hydrodynamic description is relevant for the understanding of systems such as the quark gluon plasma~(QGP) generated in heavy ion collisions. It was recently reported to be the most vortical fluid created in any experiment~\cite{STAR:2017ckg}, and is believed to possess a large angular momentum for non-central collisions. This motivates including spins, spin-orbit coupling, and spin currents in the hydrodynamic description of rotating/vortical fluids; this has been studied in a large number of publications for which we refer to the reviews~\cite{Huang:2020xyr,Florkowski:2018fap} and to~\cite{Becattini:2013fla,Xie:2015xpa,Florkowski:2017ruc}.\footnote{See~\cite{Nair:2011mk,Karabali:2014vla} for a group theoretic approach.} Before taking on that task, we here work out the effects of rotation and angular momentum on hydrodynamics, in particular on the transport coefficients measured in a rotating quantum fluid. In turn, our analysis provides the corresponding results for the quasinormal modes~(QNMs) and the transport effects associated with spinning black holes. 

Rotation and angular momentum non-trivially affect the gravitational solutions, i.e.~the metric receives off-diagonal components, and besides the mass now depends on another label, the angular momentum. 
Like their non-rotating counterparts, rotating black holes show thermodynamic behavior when for example a Hawking temperature $T$, entropy density $s$, and internal energy density $\epsilon$ are defined. These definitions, however, depend on the angular momentum now, and thermodynamic relations are modified to include angular momentum and angular 
velocity~\cite{Hawking:1998kw,Hawking:1999dp,Gibbons:2004ai}. There are now two pressures, one transverse to the angular momentum, $P_\perp$, one longitudinal to it, $P_{||}$. Furthermore, the angular velocity $\Omega$ acts as a chemical potential for the angular momentum density $\mathfrak{j}$. Hence, relations such as $\epsilon+P=s\, T$, need to be extended by a term $\Omega\, \mathfrak{j}$, and modified.  
Similarly, perturbations, quasinormal modes~(QNMs), in particular the hydrodynamic QNMs, are affected as well.  Holographically dual to that, the thermodynamics and hydrodynamics of a spinning quantum fluid are affected in a seemingly complicated way, giving the transport coefficients a dependence on the angular momentum of the fluid. Angular momentum introduces an anisotropy, which requires an anisotropic version of hydrodynamics, for examples see~\cite{Martinez:2010sc,Martinez:2010sd,Ryblewski:2010bs,Ryblewski:2011aq,Ryblewski:2012rr,Florkowski:2012lba,Strickland:2014pga,Erdmenger:2010xm,Erdmenger:2014jba,Huang:2011dc,Ammon:2017ded,Ammon:2020}. 
Similar to the two pressures, $P_\perp$ and $P_{||}$, there are now two sets of transport coefficients, those transverse and those longitudinal to the anisotropy.  
Notably, the shear viscosity splits into $\eta_\perp$ and $\eta_{||}$, and into the two corresponding shear diffusion coefficients, $\mathcal{D}_\perp$ and $\mathcal{D}_{||}$. 
In addition there are two more sets of transport coefficients, the ones measuring transport aligned or the ones measuring it anti-aligned with the angular momentum of the fluid. This yields two modified speeds of sound $v_{s,\pm}$ and two sound attenuation coefficients $\Gamma_{\pm}$.\footnote{In this work we have not considered hydrodynamic modes propagating tranverse to the angular momentum of the fluid. 
}

In this work, we provide a simple understanding of, and analytic equations for the effect of rotation on the hydrodynamic quasinormal modes of spinning black holes. Furthermore, we provide a simple understanding of, and analytic equations for the transport coefficients in the holographically dual spinning quantum fluid. 

Our results show by explicit computation that a spinning black hole or a spinning quantum fluid can both be thought of as a fluid at rest which has been boosted into a rotational motion~\cite{Kovtun:2019hdm,Hoult:2020eho}.  
In the hydrodynamic regime, we provide analytic expressions for the hydrodynamic transport coefficients as a function of the angular momentum of the black hole / fluid. We find that the hydrodynamic regime in these rotating setups is reached for perturbations with small angular momentum (equivalent to the small linear momentum limit of linear fluid motion), and, as usual, at late time (for small frequency). The equations summarizing our results are collected in the discussion Sec.~\ref{sec:discussion}, see Eqs.~\eqref{eq:etaPerpSum} to~\eqref{eq:transport0ToTransportaSound}.

\begin{figure}
    \centering
    \includegraphics[width=0.75\textwidth]{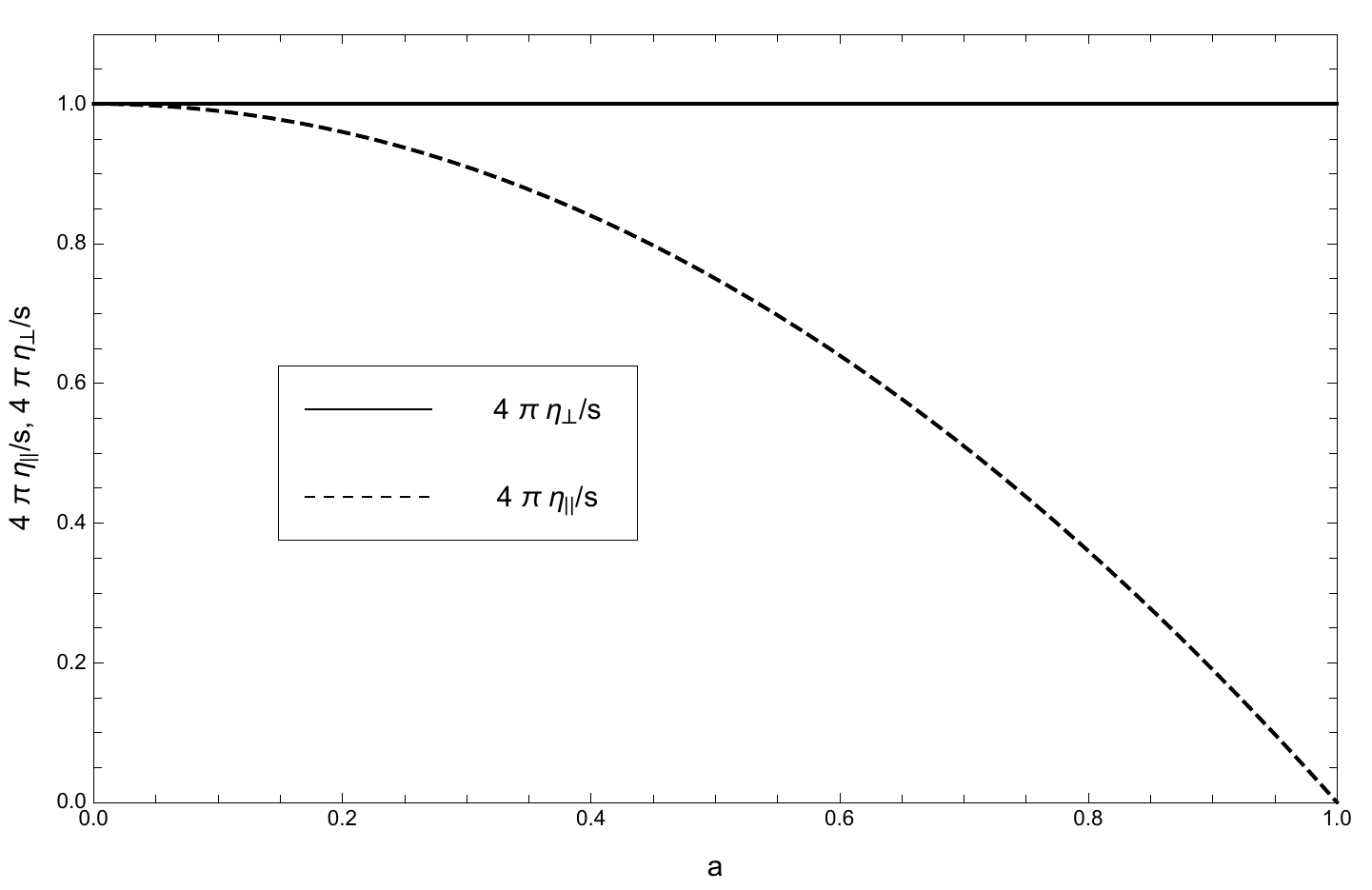}
    \caption{\label{fig:specificShearViscosities}
    {\it Ratio of shear viscosities and entropy density in a spinning black hole.} $\eta_{||}/s$ and $\eta_{\perp}/s$ are both shown as function of the rotation parameter $a$. The corresponding shear viscosity ratios for the spinning planar limit black brane are coincident with those shown. 
    }
\end{figure}

It should be stressed that the effects of rotation on measurements are not at all trivial, despite the fact that we can understand them easily. As a prominent example, consider the specific shear viscosities we have computed for a spinning quantum fluid
\begin{eqnarray} 
\label{eq:specificShearViscosityT}
 \frac{\eta_\perp}{s}&=& \frac{1}{4\pi} \, , \\
 \label{eq:specificShearViscosityL}
 \frac{\eta_{||}}{s}&=& \frac{1}{4\pi} (1-a^2)\, , \qquad -1< a < 1 \, ,
\end{eqnarray}
with the (dimensionless and normalized) angular momentum $a$; see Fig.~\ref{fig:specificShearViscosities}. Depending on the angle between the spatial direction of the measurement and the angular momentum, one will measure either  \eref{eq:specificShearViscosityT}, or \eref{eq:specificShearViscosityL}, or a combination of the two which can be parametrized by that angle \sout{$\theta$}. 
Such angle-dependent relations are conjectured for the shear diffusion mode in Sec.~\ref{sec:resultsI}, Eq.~\eqref{eq:rotatingFluidDiffusion} for arbitrary angles, and for the sound modes in two limiting cases, Eq.~\eqref{eq:rotatingFluidSoundParallel}, \eqref{eq:rotatingFluidSoundPerpendicular}. 
For fluids in {\it linear} motion the equivalent relations have been given in~\cite{Kovtun:2019hdm,Hoult:2020eho}.

For hydrodynamic simulations of heavy ion collisions (such as~\cite{Gale:2012rq,Niemi:2011ix}) this implies that different values of the specific shear viscosity have to be used, depending on the spatial direction in the QGP. It is still widely believed that at large 't Hooft coupling in all field theories with a gravity dual the shear viscosity over the entropy density has to take the value $1/(4\pi)$. This is not true\footnote{
Counter examples are provided in holographic setups involving distinct ways of breaking rotational invariance (explicitly by a magnetic field, by extra matter content, or spontaneously by a condensate) and hence inducing a spatial anisotropy~\cite{Erdmenger:2010xm,Critelli:2014kra,Ammon:2020,Rebhan:2011vd,Natsuume:2010ky}. 
In the latter setups, just like in ours, $\eta_{||}/s$ need not be $1/(4\pi)$. See a detailed discussion of this point in Sec.~\ref{sec:resultsII}. 
Of course, there are well known finite $N$ and finite 't Hooft coupling corrections violating the bound~\cite{Kats:2007mq,Brigante:2007nu,Brigante:2008gz}. 
Similar effects have been observed in systems breaking translation invariance~\cite{Ciobanu:2017fef,Burikham:2016roo,Hartnoll:2016tri} (even if isotropy was preserved).} as discussed in Sec.~\ref{sec:resultsII}.

On the gravity side, we compute all of the lowest quasinormal modes of large simply spinning five-dimensional Myers-Perry black holes as a function of the angular momentum of the black hole $a$ and the angular momentum carried by the perturbation. We show that these 
black holes in the strict large horizon limit are perturbatively stable for all angular momenta below extremality, $-L<a<L=1$ with the radius $L$ of the $AdS_5$ set to unity by rescaling as usual. We have not considered perturbations of the $S^5$ part of the ten-dimensional product space on which the type IIB supergravity (string theory) lives.

Holography was used to show that large rotating black holes in global $D$-dimensional $AdS$ spacetimes are dual to stationary solutions of the relativistic Navier-Stokes equations on $S^{D-2}$~\cite{Bhattacharyya:2007vs}. These stationary / thermodynamic  results~\cite{Bhattacharyya:2007vs} motivate an analysis of the hydrodynamic behavior of large black holes. 
Rotating black holes have been considered in the context of the AdS/CFT correspondence before, pioneered by~\cite{Hawking:1998kw,Hawking:1999dp}. QNM frequencies of rotating (charged) $AdS_4$ black strings were computed, and the hydrodynamic modes of the dual two-dimensional field theory were discussed~\cite{Morgan:2013dv,Miranda:2014vaa,Mamani:2018qzl}. Apart from the spacetime dimension, this setup differs from ours in that its 
topology is cylindrical and is locally a boosted black brane.\footnote{In contrast to that,  
we consider large black holes with spherical topology except for limiting cases discussed in detail in Sec.~\ref{sec:holographicSetup}.} 
While our perturbations (except for limiting cases) carry angular momentum, the perturbations in~\cite{Mamani:2018qzl} propagate with linear momentum. However, we can still compare the general form of the hydrodynamic shear diffusion and sound dispersion relations, to our results and to~\cite{Kovtun:2019hdm,Hoult:2020eho}. We do so in the discussion section~\ref{sec:discussion}. 
A similarity in the form of equations of motion is pointed out in Sec.~\ref{sec:resultsII}. 
Stability
has been tested, see e.g.~\cite{Murata:2007gv,Murata:2008xr,Murata:2008yx,Shibata:2009ad,Dias:2010eu,Dias:2013sdc,Cardoso:2013pza,Green:2015kur,Ganchev:2016zag,Sullivan:2017agx,Chesler:2018txn,Aminov:2020yma,Cardoso:2013pza,Horowitz:1999jd,Green:2015kur}. 
Rotating black holes have been proposed as holographic duals of the rotating QGP in~\cite{McInnes:2013wba,McInnes:2014haa,McInnes:2016dwk,McInnes:2017rxu}. While relevant phenomenological aspects are discussed, these references do not consider perturbations, QNMs, or discuss the hydrodynamic behavior of the rotating black holes. 
The drag force on a quark in a rotating QGP has been proposed in a holographic context~\cite{Arefeva:2020jvo}. 
Certain aspects of (conformal) fluids dual to rotating black holes distinct from our 
analysis have been considered~\cite{Caldarelli:1999xj,Klemm:2014nka,Berman:1999mh,Erdmenger:2014jba,Klemm:1997ea}. 
An analytic solution was found for a holographic heavy ion collision including longitudinal as well as elliptic flow, and vorticity~\cite{Bantilan:2018vjv}. 
In the context of the fluid/gravity correspondence, 
a uniformly rotating incompressible fluid~\cite{Dey:2020ogs}, the Coriolis effect~\cite{Wu:2015pzg}, 
and the diffusion constant of slowly rotating black three-branes~\cite{Amoozad:2017hrl} were studied. 
Taking the Wilsonian approach to a fluid/gravity correspondence~\cite{Bredberg:2010ky}, rotating black holes have been considered in~\cite{Lysov:2017cmc}.
\section{Rotating black holes and planar limit black branes}
\label{sec:holographicSetup}
Einstein gravity holographically corresponds to $\mathcal{N}=4$ Super-Yang-Mills~(SYM) theory, derived in a top-down construction from string theory~\cite{Maldacena:1997re}. Without destroying this correspondence, the SYM theory can be defined on a compact spacetime $S^3\times \mathbb{R}$. There are several advantages to such a formulation, for example the deconfinement transition corresponding to the Hawking-Page transition~\cite{Hawking:1982dh} can be studied, see e.g.~\cite{Klebanov:2007ws}. We consider spinning black holes, more precisely, five-dimensional Myers-Perry black holes. These have been found as vacuum solutions within Einstein gravity~\cite{Myers:1986un}; these solutions are written such that the boundary is compact, and the dual SYM lives on $S^3\times \mathbb{R}$. 
Since we are interested in a dual to a spinning fluid in flat space, $\mathbb{R}^{3,1}$, we consider large black holes. This also allows us a large temperature hydrodynamic limit later. We will distinguish three large black hole limits labeled \hyperlink{caseii}{(i)}, \hyperlink{caseii}{(ii)}, and \hyperlink{caseii}{(iii)}, as summarized in Sec.~\ref{sec:threeLimits}.

\subsection{Background metric and thermodynamics}
We investigate the consequences of rotation on the gravity side for the dual field theory. Hence, we are interested in  higher dimensional ($D>4$) rotating black holes, which are asymptotic to $AdS_{D}$. These black holes are referred to as Myers-Perry black holes in $AdS_D$ (MPAdS). Since we have applications to the quark-gluon plasma in mind, we are analyzing the five-dimensional case~(5DMPAdS), first studied in~\cite{Hawking:1998kw}, and expanded to the higher dimensions in~\cite{Gibbons:2004js,Gibbons:2004ai}.
These black holes live in global $AdS_5$, which means that the spatial coordinates on the AdS-boundary live on a three-sphere, and the topology of the dual field theory is $\mathbb{R}^{1}\times S^3$. Also the spatial coordinates on the horizon live on a three-sphere. In order to relate to fluids in flat spacetime with topology $\mathbb{R}^{3,1}$, we provide two limits. 
First, we discuss the case of large black holes, i.e.~black holes with a larger horizon radius, $r_+\gg L$, compared to the $AdS$-radius, $L$.
This will be referred to as {\it case \hyperlink{caseii}{(i)}}. As a second alternative, we perform a large black hole limit on the level of the fluctuation equations of motion, referred to as {\it case \hyperlink{caseii}{(ii)}}. As a third case, {\it case \hyperlink{caseii}{(iii)}}, we consider a large black hole limit on the level of the background metric first and then derive fluctuation equations.

For the benefit of the reader, we start with the more familiar metric as written by Hawking et al.~\cite{Hawking:1998kw} 
\begin{equation}\label{eq:metricHawkingGenAdSKerr}
    \begin{aligned}
        {ds}^2 = & - \frac{\Delta}{\rho^2}\left( dt_H - \frac{a \sin^2{\theta_H}}{\Xi_a} d\phi_H - \frac{b \cos^2{\theta_H}}{\Xi_b} d\psi_H \right)^2 + \\ 
        & \frac{\Delta_{\theta_H} \sin^2{\theta_H}}{\rho^2} \left( a dt_H - \frac{r_H^2 + a^2}{\Xi_a} d\phi_H \right)^2 + \frac{\Delta_{\theta_H} \cos^2{\theta_H}}{\rho^2} \left( b dt_H - \frac{r_H^2 + b^2}{\Xi_b} d\psi_H \right)^2 + \\
        & \frac{\rho^2}{\Delta} {dr_H}^2 + \frac{\rho^2}{\Delta_{\theta_H}} {d\theta_H}^2 + \\
        & \frac{1 + r_H^2/L^2}{r_H^2 \rho^2} \left( a b dt_H - \frac{b (r^2 + a^2) \sin^2{\theta_H}}{\Xi_a} d\phi_H - \frac{a (r^2 + b^2) \cos^2{\theta_H}}{\Xi_b} d\psi_H \right)^2 \, ,
    \end{aligned}
\end{equation}
with the definitions 
\begin{equation}
    \begin{aligned}
        \Delta = & \frac{1}{r_H^2} (r_H^2 + a^2) (r_H^2 + b^2)(1 + r_H^2/L^2) - 2 M \,, \\
        \Delta_{\theta_H} = & 1 - \frac{a^2}{L^2} \cos^2{\theta_H} - \frac{b^2}{L^2} \sin^2{\theta_H} \,, \\
        \rho = & r_H^2 + a^2 \cos^2{\theta_H} + b^2 sin^2{\theta_H} \,, \\
         \Xi_a & = 1 - a^2/L^2 \,, \\
         \Xi_b & = 1 - b^2/L^2 \, .
    \end{aligned}
\end{equation}
This metric is written in the time, $t_H$, $AdS$ radial coordinate $r_H$, and with the angular Hopf coordinates, ($\phi_H,\, \psi_H,\, \theta_H$). 

On the three-sphere, two independent angular momentum operators are necessary to generate all possible rotations, see e.g.~\cite{Hawking:1998kw}. We label the associated angular momentum parameters \{$a$, $b$\}. We will focus on the case where $a = b$. This case we refer to as the {\it simply spinning} 5DMPAdS. When $a = b$, the metric has the property of Cohomogeneity-1.
Cohomogeneity-1 is the property that the metric components can be written to only depend on one coordinate, in this case the radial $AdS$ coordinate, $r$. 
For the remainder of this work, we employ more convenient coordinates and reparameterize the mass following~\cite{Murata:2008xr} 
\begin{equation}
    \begin{aligned}
         t&=t_H \, , \\
         r^2&=\frac{a^2+r_H^2}{1-\frac{a^2}{L^2}} \, , \\
         \theta &=2 \theta_H \, ,\\
         \phi &=\phi_{H}-\psi_{H} \\
         \psi &=-\frac{2 a t_H}{L^2} +\phi_{H}+\psi_{H} \, , \\
         b &= a \, , \\
         \mu &=\frac{M}{\left(L^2-a^2\right)^3} \, .
    \end{aligned}
\end{equation}
In these new $(t,r,\theta,\phi,\psi)$ coordinates, the metric simplifies to 
\begin{equation}\label{eq:metric5dmpadssig123}
   ds^2 = -\left( 1 + \frac{r^2}{L^2} \right) {dt}^2 + \frac{{dr}^2}{G(r)} + \frac{r^2}{4} \left( (\sigma^1)^2 + (\sigma^2)^2 + (\sigma^3)^2 \right) +  \frac{2 \mu}{r^2} \left(dt + \frac{a}{2} \sigma^3 \right)^2\, ,
\end{equation}
where the blackening factor is given by 
\begin{equation}
    G(r) =  1+\frac{r^2}{L^2}-\frac{2 \mu (1-a^2/L^2)}{r^2}+\frac{2 \mu a^2}{r^4}\, , \qquad \mu = \frac{r_+^4 \left(L^2+r_+^2\right)}{2 L^2 r_+^2-2 a^2 \left(L^2+r_+^2\right)}
    \label{eq:blackeningfactor} \, ,
\end{equation}
and the spatial part of the metric is expressed in terms of three covectors 
\begin{equation}
    \begin{aligned}
        \sigma^1  & = - \sin{\psi} d\theta + \cos{\psi} \sin{\theta} d\phi \, , \\
        \sigma^2 & = \cos{\psi} d\theta + \sin{\psi} \sin{\theta} d\phi \, , \\
        \sigma^3 & = d\psi + \cos{\theta} d\phi \, ,
    \end{aligned}
    \label{eq:framesig123}
\end{equation}
with three angles $\theta,\,\psi,\, \phi$, which parameterize the three-sphere. The range of each coordinate is given by
\begin{equation}
    -\infty < t < \infty  , ~~ r_+ < r < \infty  , ~~ 0 \leqslant \theta < \pi  , ~~ 0 \leqslant \phi < 2 \pi, ~~ 0 \leqslant \psi < 4 \pi \, .
\end{equation}
As prophesied above, the $AdS$ boundary of these black holes is global $AdS_5$ with topology $S^3\times\mathbb{R}$. 

The outer horizon radius, $r_+$, is defined as the radius at which the blackening factor, $G(r_+)=0$, vanishes. The angular velocity of the outer horizon, $\Omega$, is defined as the ``$\psi/2$'' component such that $\left.\chi^2=0 \right\vert_{r=r_+}$ where $\chi=\partial_t-\Omega~\partial_{\psi/2}$ and $\Omega = \frac{a \left(L^2 + r_+^2 \right)}{L^2 ~ r_+^2}$.

According to~\cite{Murata:2008xr}, simply spinning ($a=b$) five-dimensional Myers-Perry black holes, defined  by~\eqref{eq:metric5dmpadssig123}, have two instabilities. 
First, a superradient instability has been reported, which occurs at large angular velocities $|\Omega L| > 1$. In that case, \eqref{eq:metric5dmpadssig123} is unstable against metric perturbations~\cite{Murata:2008xr}, which grow exponentially.  
This instability is not present in the regime $|\Omega L| < 1$ we consider in this work. The case $|\Omega L| = 1$ is referred to as the {\it extremal case} or the system being at {\it extremality}.
\footnote{In the strict large black hole limit ($r_+ \to \infty $), case \hyperlink{caseii}{(ii)}, $\Omega =a/L^2$, and extremality occurs at $|a| = L$.
Heuristically, this instability occurs at the extremal angular velocity due to faster than light motion at large black hole radii. 
}

A second, different instability was found~\cite{Murata:2008xr} at small horizon radius $r_+\sim L$. This is a Gregory-Laflamme~(GL) instability in the product space $AdS_5\times S^5$; this instability is not yet confirmed for the case in which only one of the two angular momenta of the black hole is nonzero ($a\neq 0,\, b=0$)~\cite{Kodama:2009bf}, or in the case of two distinct nonzero angular momenta ($a \neq b$). 
This GL instability is not within the range of parameters which we consider. In our case, the black holes are large $r_+\gg L$ and considered in the large black hole phase as discussed in~\cite{Murata:2008xr}. Below, we compute all metric perturbations and confirm that in the regime $r_+\gg L$ the 
five-dimensional Myers-Perry black hole is perturbatively stable. 
%

\subsubsection{Thermodynamics}
Thermodynamic relations have been worked out and checked regarding their consistency previously~\cite{Hawking:1998kw,Hawking:1999dp,Gibbons:2004ai}. 
As usual, the entropy density, $s$, is proportional to the ratio of the area of the black hole, $A_{\text{bh}}$, and the spatial volume of the boundary metric, $V_{\text{bdy}}$~\cite{Bardeen:1973gs,Bekenstein:1973ur}
\begin{equation}
    s=\frac{A_{\text{bh}}/(4 G_5)}{V_{\text{bdy}}}=\frac{\frac{\pi ^2 L r_+^4}{2 G_5}\sqrt{\frac{1}{L^2 r_+^2-a^2 \left(L^2+r_+^2\right)}}}{2 \pi^2 L^3}=\frac{r_+^4}{4 G_5 L^2 \sqrt{L^2 r_+^2-a^2 \left(L^2+r_+^2\right)}}\, .
\end{equation}
This is in agreement with \cite{Ishii:2018oms}. The temperature is given by 
\begin{equation}
    T = \frac{L^2 r_+^2 \left(L^2+2 r_+^2\right)-2 a^2 \left(L^2+r_+^2\right)^2}{2 \pi  \sqrt{L^8 r_+^6-a^2 L^6 r_+^4 \left(L^2+r_+^2\right)}} \, .\\
\end{equation}

According to the Ashtekar-Das Method\footnote{The Ashtekar-Das method is equivalent to the Skenderis method up to a Casimir term which is not included in the Ashtekar-Das method.
}~\cite{Kinoshita:2008dq}, the boundary stress energy tensor is given by
\begin{equation}
    \begin{aligned}
        T_{a b}\sigma^a \sigma^b &=\frac{\mu}{32 \pi  G_5 L^4}  (4 \left(a^2+3 L^2\right) {dt}^2 + L^2 \left(L^2-a^2\right) (\sigma ^1)^2+L^2 \left(L^2-a^2\right) (\sigma ^2)^2+ \\ 
        & L^2 \left(3 a^2+L^2\right) (\sigma^3)^2 +16 a L^2 dt \sigma^3 ) \, .
    \end{aligned}
\end{equation}

The angular momentum, angular velocity, energy density, pressure  of the simply spinning black hole are given by
\begin{eqnarray}\label{eq:thermoRotating}
 \mathfrak{j} &=& \frac{a \mu}{2\pi L^2 G_5}  \, , \\ 
 \Omega &=&  \frac{a \left(L^2 + r_+^2 \right)}{L^2 ~ r_+^2} \, , \\ 
 \epsilon &=& \frac{\mu (a^2+3 L^2)}{8 \pi G_5 L^4} \, , \\ 
 P_\perp &=& \frac{\mu (L^2-a^2) }{8 \pi G_5 L^4}\, , \\
 P_{||} &=& \frac{\mu (3 a^2+ L^2)}{8 \pi G_5 L^4} \, ,\\
 \mu &=& \frac{r_+^4 \left(L^2+r_+^2\right)}{2 L^2 r_+^2-2 a^2 \left(L^2+r_+^2\right)}  .
\end{eqnarray}
In the formulation of~\cite{Gibbons:2004ai}, the rotating black hole thermodynamic quantities have been shown to satisfy the thermodynamic consistency relation
\begin{equation}\label{eq:bhThermoRelation}
    \epsilon - T I_5 = s \, T + 2 \, \mathfrak{j}\,  \Omega \, ,
\end{equation}
where $\Phi = -T I_5$ is the thermodynamic potential relevant in this ensemble, and $I_5$ is the Euclidean Einstein-Hilbert action,
evaluated on the rotating black hole solution~\eqref{eq:metricHawkingGenAdSKerr},
$I_5=\frac{\pi}{4 T \Sigma_a^2}\left[ 
M-L^{-2} (r_+^2+a^2)^2
\right]$.  
The quantity $I_5$ replaces the pressure in the isotropic non-rotating version of the black hole thermodynamic  relation~\eqref{eq:bhThermoRelation}. It is clear that in an anisotropic system there are two pressures. Hence, it is not clear how each would contribute to~\eqref{eq:bhThermoRelation}. This problem was addressed in~\cite{Gibbons:2004ai} with the result given above. 
To our knowledge, the physical meaning of $I_5$ has not been clarified within holography. For anisotropy caused by a magnetic field the relations between pressures and different thermodynamic potentials is discussed in~\cite{Ammon:2012qs}, see also~\cite{Mateos:2011tv}. Since this is peripheral to the topic of our paper, we plan on returning to this question in future work.

\subsubsection{Three distinct large black hole limits} \label{sec:threeLimits}
We consider three cases:  
\paragraph{Case (i):}    \hypertarget{casei} 
             Large black holes with a large horizon value $r_+\gg L$ compared to the AdS-radius $L$. There is no limit or rescaling. The dual field theory lives on a small patch covering part of a compact spacetime $S^3\times \mathbb{R}$. QNM frequencies, $\omega$, result from a Fourier transformation of the time direction. They are reported as functions of the discrete total angular momentum parameter $\mathcal{J}$ of the relevant metric perturbation. Through a rescaling of parameters and coordinates it was shown that in this case QNM frequencies of the non-rotating global AdS are asymptotic to planar black brane QNM frequencies~\cite{Horowitz:1999jd}.
    \paragraph{Case (ii):} \hypertarget{caseii} 
             Large black holes with the strict limit $r_+ \to \infty$ performed in the fluctuation equations which determine the QNMs (after deriving them). There is no limit performed on the Myers-Perry black hole metric, i.e.~the geometry is that of a spinning black hole, not that of a black brane. The dual field theory lives on a small patch covering part of a compact spacetime $S^3\times \mathbb{R}$. This case involves a rescaling of the frequency 
             \begin{equation}
                \omega\to\alpha 2\nu r_+/L \, , \quad \alpha\to\infty
             \end{equation}
             and the angular momentum of the fluctuation 
             \begin{equation}
                \mathcal{J}\to\alpha j r_+/L\, , \quad \alpha\to\infty
             \end{equation} 
             as well as the black hole horizon radius
             and radial coordinate
             \begin{equation}
                 r_+\to \alpha r_+\, , \quad \alpha\to\infty \, ,
             \end{equation}
             \begin{equation}
                 r\to \alpha r\, , \quad \alpha\to\infty \, ,
             \end{equation}
             and leading order terms in $\alpha$ are kept in the fluctuation equations. QNM frequencies are reported in terms of the dimensionless frequency
            $\nu$, which is a function of the now continuous dimensionless angular momentum parameter $j$.
In this limit, the thermodynamic quantities, Eq.~\eqref{eq:thermoRotating} asymptote to 
\begin{eqnarray}
 s&\sim& \frac{r_+^3}{4 G_5 L^2 \sqrt{L^2-a^2}} \, , \\
 \mathfrak{j} &\sim& \frac{a \mu}{2\pi L^2 G_5}  \, , \\ 
 \Omega &\sim&  \frac{a}{L^2} \, , \\ 
 \epsilon &\sim& \frac{\mu (a^2+3 L^2)}{8 \pi G_5 L^4} \, , \\ 
 P_\perp &\sim& \frac{\mu (L^2-a^2) }{8 \pi G_5 L^4}\, , \\
 P_{||} &\sim& \frac{\mu (3 a^2+ L^2)}{8 \pi G_5 L^4} \, , \\
 \mu &\sim& \frac{r_+^4}{2 (L^2 - a^2)} \, ,\\
 T &\sim& \frac{\sqrt{L^2-a^2}~r_+}{\pi L^3} \, .
\end{eqnarray}
    
    \paragraph{Case (iii):} \hypertarget{caseiii} 
            Planar limit black brane as a limit of the black hole. The limit $r_+\to \infty$ is performed in the black hole metric turning it into the metric of a boosted black brane. 
            We stress that most of our results are not obtained in this limit, but in the other cases \hyperlink{caseii}{(i)} and \hyperlink{caseii}{(ii)}. The planar limit 
            (iii) is merely used for comparison to the literature and as a more familiar setup\footnote{More familiar at least to us.}. 
            In more detail, start with the black hole metric, Eq.~\eqref{eq:metric5dmpadssigpm3}. 
            First, apply the coordinate transformation
            \begin{equation}\label{eq:simpcoordtranstoxyz}
            \begin{aligned}
        t  & = \tau \, ,\\
        \frac{L}{2} (\phi - \pi)  & = \text{$x$}  \, , \\
        \frac{L}{2} \tan \left( \theta-\pi/2 \right)  & = \text{$y$} \, ,\\
        \frac{L}{2} (\psi-2 \pi)  & = \text{$z$} \, , \\
        r & = \tilde{r} \, ,
    \end{aligned}
    \end{equation}
    where $\tau$, $\tilde{r}$, $x$, $y$, and $z$ are the new coordinates. In a second step, we scale each of the coordinates with an appropriate power of a scaling factor $\alpha$
    \begin{equation}\label{eq:simpalphalargeTscale}
    \begin{aligned}
        \tau  & \rightarrow \alpha^{-1} \tau \, ,\\
        x  & \rightarrow \alpha^{-1} x \, ,\\
        y  & \rightarrow \alpha^{-1} y \, ,\\
        z  & \rightarrow \alpha^{-1} z \, ,\\
        \tilde{r} & \rightarrow \alpha ~ \tilde{r} \, ,\\
        r_+ & \rightarrow \alpha ~ \tilde{r}_+\, , \quad (\alpha \to \infty) \, .
    \end{aligned}
    \end{equation}
    This leads to a Schwarzschild black brane metric  
    that has been boosted about the $\tau$-$z$ plane with boost parameter $a$,
\begin{equation}\label{eq:metricBoostedBlackBrane}
    ds^2_{\text{brane}} = \frac{r^2}{L^2} \left( -{d\tau}^2 + d{x}^2+ d{y}^2+ d{z}^2 + \frac{r_+^4}{r^4 \left( 1-a^2/L^2 \right)} \left( d\tau + \frac{a}{L} d{z} \right)^2 \right)+\frac{L^2 r^2}{r^4-r_+^4} {dr}^2 \, .
\end{equation}
The Schwarzschild black brane metric is recovered by the boost
\begin{equation} \label{eq:boostTrafo}
    \begin{aligned}
       \tau =& \frac{L}{\sqrt{L^2-a^2}} t'+\frac{a}{\sqrt{L^2-a^2}}(z)'\, ,\\
        z =& \frac{a }{\sqrt{L^2-a^2}} t'+\frac{L}{\sqrt{L^2-a^2}}(z)' \, ,\\
       \tilde{r} =& r' \, ,\\ 
        x =& (x)' \, ,\\ 
        y =& (y)'\, .
    \end{aligned}
\end{equation}
The entropy density and temperature of the boosted black brane  are given by
\begin{equation}\label{eq:thermodynamicsBoostedBlackBrane}
    s = \frac{r_+^3}{4 G_5 L^2 \sqrt{L^2-a^2}}\, ,  \quad T= \frac{r_+ \sqrt{L^2-a^2}}{\pi  L^3} \, .
\end{equation}
   This limit we refer to as a {planar limit} because it ``zooms in'' on a flat Poincare patch covering a small part of the large black hole. This turns the three compact spatial coordinates ($\theta,\, \psi, \, \phi$) on the $S^3$ into non-compact coordinates ($x,\, y,\, z$)$\in [-\infty,\infty]$. In other words, this turns global $AdS$ into Poincare patch $AdS$.

    \paragraph{Comments on the planar limit black brane, case (iii):} 
    While it is not our focus limit, there are two motivations for the limit \hyperlink{caseii}{(iii)}. First, we intend to move to a regime of large temperature in order to have a dual field theory on a {\it non-compact spacetime}, because this is more relevant for applications to heavy ion collisions or condensed matter systems. Second, we know that at large temperature, a Schwarzschild black brane will show {\it hydrodynamic behavior} if the gradients are small, providing a setup for comparison which is well-studied at $a=0$. 
    
    For the metric  \eqref{eq:metric5dmpadssigpm3}, a large temperature limit is admitted if $|a|<L$. 
    A large horizon radius, $r_+$ corresponds to a large temperature. Naively one would think of only scaling $r$ and $r_+$ to achieve a large temperature black hole but in order to have a well defined scaling limit one must also scale the non-radial coordinates, naturally leading to a black brane.
    The dual field theory lives in the resulting flat spacetime $\mathbb{R}^{3,1}$. The fluctuation equations are derived within that flat black brane metric. QNMs are reported in terms of the frequency $\omega$ and the linear momentum $k$, which follow from a Fourier transformation of the time and the spatial coordinate, $z$, which is aligned with the angular momentum.  
    
    This case is different from case \hyperlink{caseii}{(ii)} because the scalings are not identical. In particular, the radial AdS coordinate $r$ is scaled in \hyperlink{caseii}{(iii)} to become large, but not scaled in \hyperlink{caseii}{(ii)}. Case \hyperlink{caseii}{(ii)} is different from case \hyperlink{caseii}{(iii)} in that it preserves the $S^3\times \mathbb{R}$ structure whilst case \hyperlink{caseii}{(iii)} is an approximation of the large black hole at large horizon radius with a black brane which has the structure $\mathbb{R}^{3,1}$ at its boundary. The quasinormal modes found in case \hyperlink{caseii}{(ii)} are all stable, those in case \hyperlink{caseii}{(iii)} become unstable at large angular momentum {\it below} extremality, hinting that case \hyperlink{caseii}{(iii)} may have maladies in its method.
    
    So, we expect all transport coefficients of this boosted black brane to assume the values of a Schwarzschild black brane at $a=0$. Indeed, this is the case for the shear viscosities and the longitudinal momentum diffusion constant. At nonzero $a$, we report these three transport coefficients as a function of the boost parameter $a$ in Sec.~\ref{sec:viscositiesPlanarLimit} and~\ref{sec:planarLimit}.
        %

\subsection{Perturbations}
We analyze the three possible channels: shear (tensor), momentum diffusion (vector), and sound propagation (scalar). 
In systems with spherical symmetry, the terms tensor, vector, and scalar are referring to how the perturbations transform under rotations. We will use these same names with a slight abuse of notation, as explained below. 
Consider the perturbed metric
\begin{equation}\label{eq:pertgeneric}
    \begin{aligned}
        g^{p}_{\mu\nu} {dx}^\mu {dx}^\nu = (g_{\mu\nu}+\epsilon~h_{\mu\nu}+O(\epsilon^2)) {dx}^\mu {dx}^\nu\, ,
    \end{aligned}
\end{equation}
where $g$ is the spinning black hole metric, which satisfies the Einstein equations (at order $\epsilon^0$). The perturbation, $h_{\mu\nu}$, now is required to satisfy the Einstein equations resulting at order $\epsilon^1$, namely
\begin{equation}\label{eq:pertgenericeom}
    \begin{aligned}
        \dot{R}_{\mu\nu} = \frac{2\Lambda}{D-2}h_{\mu\nu}\, ,
    \end{aligned}
\end{equation}
with the dot denoting the derivative with respect to the parameter $\epsilon$, the cosmological constant $\Lambda$ and $$\dot{R}_{\mu\nu} = -\frac{1}{2}\nabla_\mu \nabla_\nu h-\frac{1}{2}\nabla^\lambda \nabla_\lambda h_{\mu\nu}+\nabla^\lambda \nabla_{(\mu}h_{\nu)\lambda}\,,$$ where $h=h^{~~\mu}_{\mu}=h_{\nu \mu} g^{\mu \nu}$.  The diffeomorphism-covariant derivatives are defined with respect to the background metric. 

\subsubsection{Symmetry}
We use the symmetry of the spacetime \eqref{eq:metric5dmpadssig123} to simplify the equations of motion of linear perturbation to ordinary differential equations. 
The symmetry group of \eqref{eq:metric5dmpadssig123} is $\mathbb{R}_t \times SU(2) \times U(1)$~\cite{Murata:2008xr}. This symmetry group is generated by five Killing vector fields. For the $\mathfrak{su}(2)$ subalgebra
\begin{equation}
    \begin{aligned}
        \xi_x & = \cos{\phi} \partial_\theta + \frac{\sin{\phi}}{\sin{\theta}} \partial_\psi - \cot{\theta} \sin{\phi} \partial_\phi\, ,\\
        \xi_y & = -\sin{\phi} \partial_\theta + \frac{\cos{\phi}}{\sin{\theta}} \partial_\psi - \cot{\theta} \cos{\phi} \partial_\phi\, ,\\
        \xi_z & = \partial_\phi \, .
    \end{aligned}
    \label{eq:kvfsu2}
\end{equation}
The $\mathfrak{u}(1)$ subalgebra is generated by $e_3 = \partial_\psi$ where $e_a$ are the 
dual vectors to $\sigma^a$. 

The time translations are 
generated by $\partial_t$. We may construct an algebra with the operators, $W_a = i e_a$ and $L_\alpha = i \xi_\alpha$.\footnote{$\alpha, \beta,... = x,y,z$ and $a,b,\dotsc = 1,2,3$.} The Lie brackets are
\begin{equation}
    [L_\alpha,L_\beta] = i \epsilon_{\alpha \beta \gamma} L_\gamma ~, ~~~ [W_a,W_b] = -i \epsilon_{abc} W_c ~, ~~~ [W_a,L_\alpha] = 0 \, .
    \label{eq:WLalgebra}
\end{equation}

For the compatible operators, $L^2$, $L_z$, and $W_3$, their
simultaneous eigenfunctions are the Wigner-D functions, $D^\mathcal{J}_{\mathcal{KM}}(\theta,\phi,\psi)$~\cite{Ishii:2020muv,certik:2009,weisstein:2020}. 
We have the following eigenvalue equations: 
\begin{equation}\label{eq:wigeigenequations}
    W^2 D^\mathcal{J}_{\mathcal{KM}}=L^2 D^\mathcal{J}_{\mathcal{KM}}=\mathcal{J}(\mathcal{J}+1)D^\mathcal{J}_{\mathcal{KM}},\quad
    L_z D^\mathcal{J}_{\mathcal{KM}}=\mathcal{M} D^\mathcal{J}_{\mathcal{KM}},\quad 
    W_3 D^\mathcal{J}_{\mathcal{KM}}=\mathcal{K} D^\mathcal{J}_{\mathcal{KM}}\, .
\end{equation}
The Wigner-D functions form a complete set on $S^3$. Heuristically, they are ``spherical harmonics'' on the $S^3$. Thus, they are related to familiar angular momentum algebra representations. In that sense one may label $\mathcal{J}$ the total angular momentum. However, since there are two angular momenta on the compact three-sphere, this representation is slightly more involved than we are used to from quantum mechanics. Let us gather a better intuition considering special cases. 
For example, they reduce to the spherical harmonics $Y_\ell^m$ (times a normalization factor) in the case $\mathcal{J}=\ell$, $\mathcal{K}=m$, and $\mathcal{M}=0$:
\begin{equation}
    \mathcal{D}^\ell_{m\, 0}(\alpha,\beta,\gamma) = \sqrt{\frac{4\pi}{2\ell+1}}(Y_\ell^m)^*(\beta,\gamma) \, .
\end{equation}
In this case, they are also proportional to the Legendre polynomials $P_\ell^m(\cos (\beta)\,) e^{-i\, m\, \alpha}$. In the limit of large total angular momentum $\mathcal{J}\gg \mathcal{K\, M}$, they are related to exponentials and the Bessel function, $J_{M M'}$, by
\begin{equation}
    \mathcal{D}^\mathcal{J}_\mathcal{KM} \approx e^{-i\, \mathcal{K} \,\alpha - i\, \mathcal{M\, \gamma}} J_{\mathcal{K-M}}(\mathcal{J}\beta) \, .
\end{equation}

For the purpose of perturbation decomposition, it was found that linear perturbations of different $\mathcal{J}$ and $\mathcal{K}$ have to obey decoupled equations of motion~\cite{Hu:1974hh,Murata:2007gv,Kimura:2007cr,Murata:2008yx}.
In order to use the results of \cite{Murata:2008xr,Murata:2007gv} for our fluctuation decomposition below, we express the metric (\ref{eq:metric5dmpadssig123}) in a more tractable form by using the orthonormal frame covectors, $\sigma^\pm = \frac{1}{2} \left(\sigma^1 \mp i \sigma^2 \right)$. The dual basis 
$e_\pm  = e_1 \pm i e_2$. In this basis \eqref{eq:metric5dmpadssig123} takes the form
\begin{equation}\label{eq:metric5dmpadssigpm3}
    \begin{aligned}
    {ds}^2 = & -\left( 1 + \frac{r^2}{L^2} \right) {dt}^2 + \frac{{dr}^2}{G(r)} + \frac{r^2}{4} \left(4 \sigma^{+}\sigma^{-} + (\sigma^3)^2 \right) +  \frac{2 \mu}{r^2} \left(dt + \frac{a}{2} \sigma^3 \right)^2\, .
    \end{aligned}
\end{equation} 

We define raising and lowering operators, $W_{\pm}:=W_1\pm i W_2$. Using (\ref{eq:wigeigenequations}), we can express the action of $W_\pm$ and $W_3$ on $D^\mathcal{J}_{\mathcal{KM}}$. We find 
$$
    \begin{aligned}
        W_+D^\mathcal{J}_{\mathcal{KM}}&=i\sqrt{(\mathcal{J}+\mathcal{K})(\mathcal{J}-\mathcal{K}+1)}D^\mathcal{J}_{\mathcal{K}-1\ \mathcal{M}} \, , \\  W_-D^\mathcal{J}_{\mathcal{KM}}&=-i\sqrt{(\mathcal{J}-\mathcal{K})(\mathcal{J}+\mathcal{K}+1)}D^\mathcal{J}_{\mathcal{K}+1\ \mathcal{M}} \, , \\ 
        W_3 D^\mathcal{J}_{\mathcal{KM}}&=\mathcal{K} D^\mathcal{J}_{\mathcal{KM}}  \, .
    \end{aligned}
$$
In terms of partials, $\partial_\pm$ and $\partial_3$, we have the more useful expressions
$$
    \begin{aligned}
        \partial_+D^\mathcal{J}_{\mathcal{KM}}&=\sqrt{(\mathcal{J}+\mathcal{K})(\mathcal{J}-\mathcal{K}+1)}D^\mathcal{J}_{\mathcal{K}-1\ \mathcal{M}} \, ,\\  \partial_-D^\mathcal{J}_{\mathcal{KM}}&=-\sqrt{(\mathcal{J}-\mathcal{K})(\mathcal{J}+\mathcal{K}+1)}D^\mathcal{J}_{\mathcal{K}+1\ \mathcal{M}} \, ,\\ 
        \partial_3 D^\mathcal{J}_{\mathcal{KM}}&=-i\mathcal{K} D^\mathcal{J}_{\mathcal{KM}}  \, ,
    \end{aligned}
$$
where $\partial_\pm={e_\pm}^\mu\partial_\mu$ and $\partial_3={e_3}^\mu\partial_\mu$.

\subsubsection{Perturbation Equations}
The Wigner-D functions form a complete set on $S^3$ and are spatial eigenmodes of the system. Likewise the exponential Fourier modes, $e^{-i\omega t}$, are the eigenmodes in time direction. So one can construct a general perturbation,
\begin{equation}\label{eq:pertsimplygeneric}
    h_{\mu\nu} = \int d\omega e^{-i\omega t} \sum_{\mathcal{J} = 0} \sum_{\mathcal{M}=\mathcal{J}}^{\mathcal{J}} \sum_{\mathcal{K}=-(\mathcal{J}+2)}^{\mathcal{J}+2} h_{a b}(r;\omega, \mathcal{(J,M),K}) \sigma^a_{\mu} \sigma^b_{\nu} D_{\mathcal{K}-Q(\sigma^{a})-Q(\sigma^{b})}^\mathcal{J} \, ,
\end{equation}
where the sum over $a$ and $b$ is implied. 
Here, $\mathcal{M}$ is dropped from the Wigner-D functions because $\mathcal{M}$ does not appear in the equations of motion, as we have confirmed by explicit computation. 
Here we defined the parameter 
\begin{equation}
    Q(\sigma^a) := \left\{ \begin{array}{cc} 0 & a=r,t,3 \\1 & a=+ \\ -1 & a=- \end{array} \right. 
\end{equation}
This ``twisted sum'' guarantees that terms of different $\mathcal{K}$ charges are added. One can see that this is the case because 
\begin{eqnarray}
    W_3 (\sigma^a_{\mu} \sigma^b_{\nu} D_{\mathcal{K}-Q(\sigma^{a})-Q(\sigma^{b})}^\mathcal{J})&=&
(W_3(\sigma^a_{\mu})) \sigma^b_{\nu} D_{\mathcal{K}-Q(\sigma^{a})-Q(\sigma^{b})}^\mathcal{J}+
\sigma^a_{\mu} (W_3(\sigma^b_{\mu})) D_{\mathcal{K}-Q(\sigma^{a})-Q(\sigma^{b})}^\mathcal{J}\nonumber\\
&& +\sigma^a_{\mu} \sigma^b_{\nu} (W_3(D_{\mathcal{K}-Q(\sigma^{a})-Q(\sigma^{b})}^\mathcal{J}))=\mathcal{K}~ \sigma^a_{\mu} \sigma^b_{\nu} D_{\mathcal{K}-Q(\sigma^{a})-Q(\sigma^{b})}^\mathcal{J} \, .
\end{eqnarray}

One can simplify the notation to
\begin{equation}
    h_{\mu\nu}^\mathcal{(K)}=\sum_{\mathcal{K}} h_{\mu\nu}^\mathcal{(K)}(x^\lambda) = \sum_{\mathcal{K}=-(\mathcal{J}+2)}^{\mathcal{J}+2} h_{a b}^\mathcal{(K)} \sigma^a_{\mu} \sigma^b_{\nu} D_{\mathcal{K}-Q(\sigma^{a})-Q(\sigma^{b})}^\mathcal{J}\, ,
\end{equation}
where perturbations of different $\mathcal{K}$ charges and different ($\mathcal{J}$,$\mathcal{M}$) decouple in \eqref{eq:pertgenericeom}. 

\paragraph{Tensor perturbations:} 
The mode $h_{++}$ decouples from all other perturbations when restricting the set of perturbations such that $\mathcal{K}=\mathcal{J}+2$. We will refer to it as the {\it tensor mode}, and to this sector of perturbations as {\it tensor sector}. Reading off the perturbations from
Eq.~\eqref{eq:pertsimplygeneric}
\begin{equation}\label{eq:pertsimplyglobaltensor}
    h^T_{\mu\nu} \equiv e^{-i\omega t} r^2 h_{++}(r) \sigma^+_{\mu} \sigma^+_{\nu} D_{\mathcal{J} \mathcal{M}}^\mathcal{J} 
    \, .
\end{equation}
Note that there is another mode, $h_{--}$, which also decouples from all other perturbations. It is the complex conjugate of $h_{++}$. $h_{++}$ satisfies a single decoupled EOM, similar to the perturbation $h_{x y}$ around a black brane (and its complex conjugate), which transforms as a tensor under $O(2)$-rotations around the momentum $k_{z}$, see e.g.~\cite{Kovtun:2005ev}.

Plugging Eq.~\eqref{eq:pertsimplyglobaltensor} into Eq.~\eqref{eq:pertgenericeom}, we obtain the equation of motion for $h_{++}$
\begin{equation}\label{eq:pertsimplyglobalpoincaretensoreom}
    \begin{aligned}
        0 = & h_{++}''(r) + \frac{\left(L^2 \left(3 r^4-2 \mu  \left(a^2+r^2\right)\right)+2 a^2 \mu  r^2+5 r^6\right)}{L^2 \left(2 a^2 \mu  r-2 \mu  r^3+r^5\right)+2 a^2 \mu  r^3+r^7} h_{++}'(r) + \\ 
        & \frac{h_{++}(r)}{\left(L^2 \left(2 a^2 \mu -2 \mu  r^2+r^4\right)+2 a^2 \mu  r^2+r^6\right)^2} \times \\
        & \quad L^2 r^2 (-4 (\mathcal{J}+2) \left((\mathcal{J}+1) r^4 \left(L^2+r^2\right)-2 \mu  \left(a^2 \left(L^2+r^2\right)+(\mathcal{J}+1) L^2 r^2\right)\right)+ \\
        & \quad L^2 r^2 \omega ^2 \left(2 a^2 \mu +r^4\right)-8 a (\mathcal{J}+2) \mu  L^2 r^2 \omega )\, .
    \end{aligned}
\end{equation} 
Note, this equation is written in ingoing  Eddington-Finkelstein coordinates after the transformation~\eqref{eq:framesimplyedfinkelstein}.

\paragraph{Vector perturbations:} 
We refer to the perturbation sector defined by  $\mathcal{K}=\mathcal{J}+1$ as the {\it vector sector}, and to the modes associated with $h_{3+},\, h_{t+}, \, h_{++}$ as {\it vector modes} where we use the radial gauge $h_{r+} = 0$. These can be written
\begin{equation}\label{eq:pertsimplyglobalvector}
    \begin{aligned}
        h^V_{\mu\nu} \equiv & e^{-i\omega t} r^2  (h_{++}(r) \sigma^+_{\mu} \sigma^+_{\nu} D_{(\mathcal{J}-1)\mathcal{M}}^\mathcal{J} + \\
        & 2 \left(h_{+r}(r) \sigma^+_{(\mu} \sigma^r_{\nu)} + h_{+t}(r) \sigma^+_{(\mu} \sigma^t_{\nu)} + h_{+3}(r) \sigma^+_{(\mu} \sigma^3_{\nu)} \right) D_{\mathcal{J}\mathcal{M}}^\mathcal{J}) \,
    \end{aligned}\text{.}
\end{equation}
Plugging the perturbations~\eqref{eq:pertsimplyglobalvector} into the perturbation equations~\eqref{eq:pertgenericeom}, yields the coupled set of equations of motion for ($h_{+t}$,$h_{+3}$,$h_{++}$), which is reported in the appendix due to excessive length, see Eq.~\eqref{eq:pertsimplyglobalpoincarevectoreom}. We have further found that~$h_{++}$ decouples from this sector in the limits of large black holes referred to as cases \hyperlink{caseii}{(ii)} and \hyperlink{caseii}{(iii)} in Sec.~\ref{sec:holographicSetup}. There are complex conjugate fluctuations $(h_{-t},\, h_{-3},\, h_{--})$, which satisfy an identical set of equations. This gives a hint that these fluctuations may be similar to the black brane fluctuations which transform like vectors under $O(2)$-rotations around the momentum $k$, see e.g.~\cite{Kovtun:2005ev}, $h_{\tau x},\, h_{z x}$, which satisfy the same eoms as $h_{x z},\, h_{y z}$.
\paragraph{Scalar perturbations:} 
We refer to the perturbation sector defined by  $\mathcal{K}=\mathcal{J}$ as {\it sound sector}, and the modes associated with it are referred to as {\it scalar modes}, where we use radial gauge again $h_{ri} = h_{rt} = h_{rr} = 0$,  
and obtain
\begin{equation}\label{eq:pertsimplyglobalscalar}
    \begin{aligned}
        h^S_{\mu\nu} & \equiv  e^{-i\omega t} r^2 ( h_{++}(r) \sigma^+_{\mu} \sigma^+_{\nu} D_{(\mathcal{J}-2)\mathcal{M}}^\mathcal{J} + \\
        & 2 (h_{+r}(r) \sigma^+_{(\mu} \sigma^r_{\nu)} + h_{+t}(r) \sigma^+_{(\mu} \sigma^t_{\nu)} + h_{+3}(r) \sigma^+_{(\mu} \sigma^3_{\nu)} D_{(\mathcal{J}-1)\mathcal{M}}^\mathcal{J} ) + \\
        & (2 h_{+-}(r) \sigma^+_{(\mu} \sigma^-_{\nu)} + \sum_{i,j\in \{t,3,r\}} h_{ij}(r) \sigma^i_{(\mu} \sigma^j_{\nu)} ) D_{\mathcal{J}\mathcal{M}}^\mathcal{J}) \, .
    \end{aligned}
\end{equation}
Plugging the perturbations~\eqref{eq:pertsimplyglobalscalar} into the perturbation equations~\eqref{eq:pertgenericeom}, yields a coupled set of equations of motion, which is not reported in this work due to mind-bogglingly excessive length.

In summary, on the background of simply spinning Myers-Perry $AdS_5$ black holes there exist three sectors of perturbations (tensor, vector, scalar), which decouple from each other when represented in terms of Wigner-D functions, $\mathcal{D}^\mathcal{J}_{\mathcal{K M}}(\theta,\phi,\psi)$, and taking the limit of large black holes.  We consider the three distinct large black hole limits \hyperlink{caseii}{(i)}, \hyperlink{caseii}{(ii)}, and \hyperlink{caseii}{(iii)}.


\section{Results I: Hydrodynamic behavior in quasinormal modes (QNMs)}
\label{sec:resultsI}
In this section we present and carefully analyze the QNMs in the three different perturbation sectors, i.e. the tensor, momentum diffusion (vector), and sound propagation (scalar) sectors. As indicated by suggestive naming of the sectors, we discover hydrodynamic behavior in two of these sectors. We begin with numerical results and then derive analytic results, which will be used as cross-check for the numerics. 

As a brief prelude, we recall what is known about hydrodynamic behavior of black holes in the non-rotating case. In non-rotating black holes it is known from~\cite{Horowitz:1999jd}  that the spectrum of (tensor) QNMs of a global AdS black hole with the horizon topology $S^3\times \mathbb{R}$ asymptotes to the QNM spectrum in the limit of large black hole radius. Our planar black brane limit, case \hyperlink{caseii}{(iii)}, is such a large black hole radius limit. Hence, our planar limit QNMs at vanishing rotation are guaranteed to display hydrodynamic behavior at large horizon radius and correspondingly large temperature, as known from previous studies~\cite{Kovtun:2005ev}. The more interesting questions are: 
\begin{enumerate}
    \item Do hydrodynamic QNMs arise generically in large rotating black holes (without taking the planar limit)?
    \item Does fast rotation destroy such hydrodynamic behavior?
\end{enumerate}
Answers eagerly await the reader below.

\subsection{QNMs of simply spinning black hole}
%
\subsubsection{Tensor sector}
As mentioned before, the fluctuation $h_{++}$ decouples from all others for the choice $\mathcal{K}=\mathcal{J}+2$, hence we refer to it as tensor fluctuation. We choose this name in analogy to the tensor fluctuation on a black brane~\cite{Kovtun:2005ev}, e.g.~$h_{x y}$, which transforms as a tensor under $SO(2)$-rotations in the plane perpendicular to the momentum, e.g. in $z$-direction with $h_{x y}(\tau,z)\propto e^{-i\omega \tau + i k z}$.  Our results show that indeed $h_{++}$ displays properties which are similar to $h_{x y}$.

\begin{figure}
    \centering
    \includegraphics[width=1\textwidth]{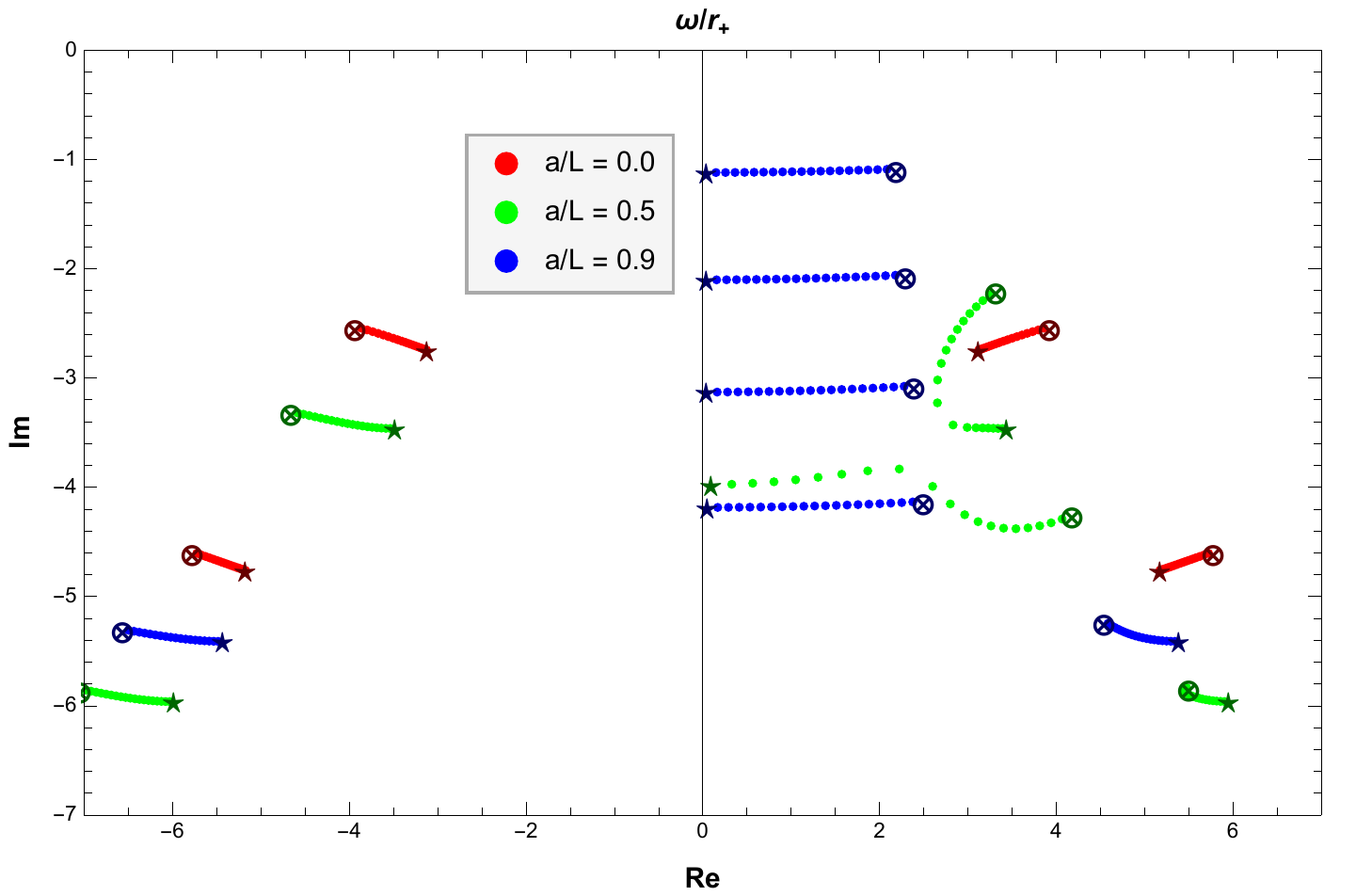}
    \caption{\label{fig:hppQNMsofJ}
    {\it Tensor sector QNMs.} The $h_{++}$ QNMs at $\mathcal{K}=\mathcal{J}+2$ for $r_+ = 100$  at $\mathcal{J}={0,5,10,20,\dots,100}$, where the star indicates the value at $\mathcal{J}=0$, the circled cross indicates $\mathcal{J}=100$. This is {{case~\protect\hyperlink{caseii}{(i)}}}.}
\end{figure}

The lowest tensor QNMs are displayed in Fig.~\ref{fig:hppQNMsofJ} for a large value of the black hole horizon $r_+=100$ in units of the AdS radius $L$
; this is case \hyperlink{caseii}{(i)} introduced in Sec.~\ref{sec:holographicSetup}. First, consider the nonrotating case, $a=0$, indicated by the red dots. When the Wigner-parameter $\mathcal{J}$ vanishes, the red stars align along two diagonals symmetric about the imaginary axis. We have checked, that this is also the case for the higher QNMs not displayed here. 
As we increase $\mathcal{J}$ from 0~(star) to 100~(cross), the QNMs all move symmetrically away from the imaginary axis and up. 
Note, that this behavior closely resembles the behavior of tensor QNMs of black branes, when displayed as function of momentum, see~\cite{Janiszewski:2015ura}.

Now we increase the angular momentum parameter to 50\% of its extremal value, $a/L=0.5$, shown in green on Fig.~\ref{fig:hppQNMsofJ}. Looking first at the stars, $\mathcal{J}=0$, we note the symmetry of the QNMs about the imaginary axis persists, as the lowest two QNMs are located approximately at $(\pm 3.75,\,-3.40)$. However, when increasing $\mathcal{J}>0$, an asymmetry about the imaginary axis becomes apparent. In the rotating case, QNMs of increasing angular momentum parameter $\mathcal{J}$ become increasingly asymmetric. This trend is confirmed by the blue data obtained at 90\% of the extremal angular momentum parameter. 
This happens since the angular momentum of the fluid / black hole breaks parity and separates modes into those which are rotating with and those rotating against the black hole / fluid. 

Interestingly, at increasing $a$, and at $\mathcal{J}=0$, there is a set of purely imaginary modes moving up the imaginary axis towards the origin. This closely resembles the behavior of tensor modes on a (charged) Reissner-Nordstr\"om black brane when increasing the charge from zero towards extremality~\cite{Janiszewski:2015ura}. At nonzero $\mathcal{J}$, however, we observe an asymmetry again, i.e.~these $h_{++}$ QNMs acquire a positive real part, being pushed to the right, away from the imaginary axis. 
\begin{figure}
    \centering
    \includegraphics[width=1\textwidth]{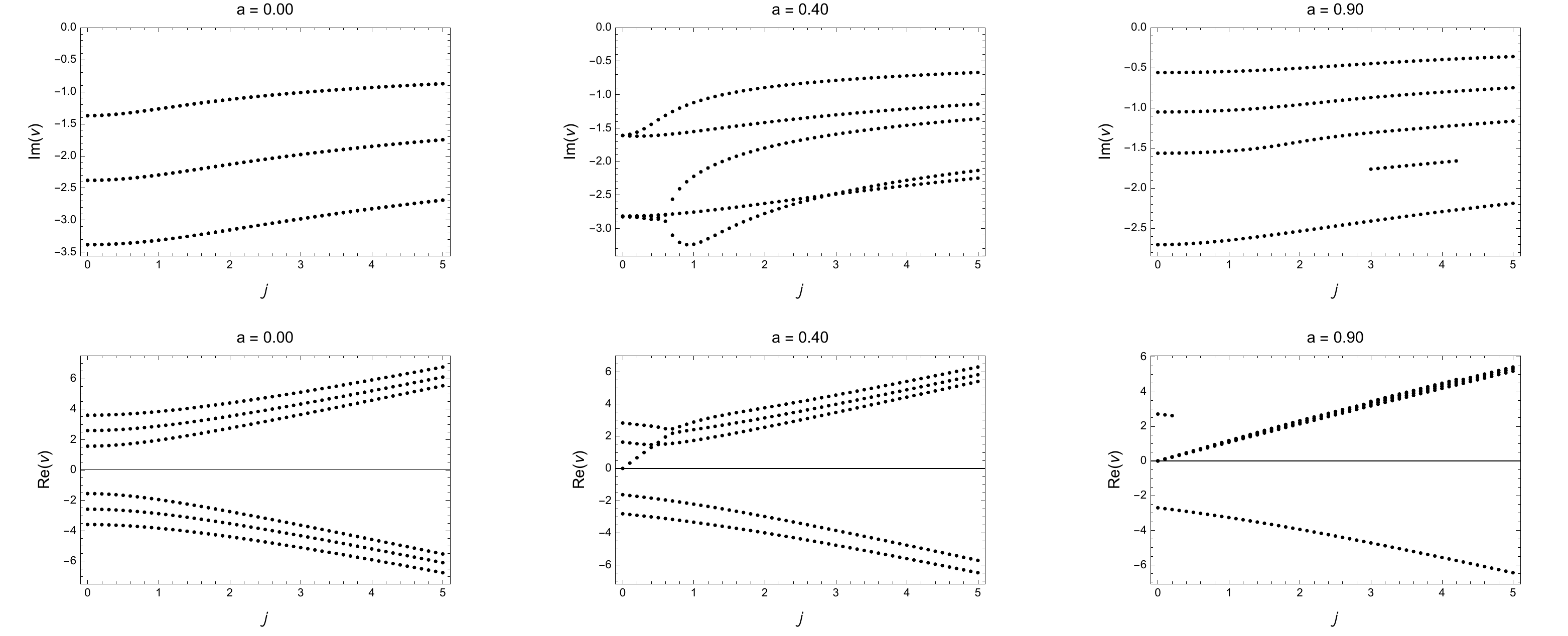}
    \caption{ \label{fig:qnfsh++}
    {\it Tensor sector dispersion relations.} Displayed are the real (lower plots) and imaginary (upper plots) part of the QNM frequencies of the $h_{++}$-perturbation around a large black hole as a function of the parameter $j$. This parameter is defined for large black holes where $\omega \rightarrow \alpha \nu$ and $\mathcal{J} \rightarrow \alpha j$ as $\alpha\to\infty$. Due to this limit on the level of the fluctuation equation for~$h_{++}$, this is {{case~\protect\hyperlink{caseii}{(ii)})}}.
    Modes displayed here were found with precision of at least $d\nu = \nu 10^{-5}$. Missing modes did not converge to sufficient precision and have been filtered out by the numerical routine.
    }
\end{figure}

In~\cite{Horowitz:1999jd} it was shown through a rescaling of parameters in the equations of motion that the QNMs of large black holes in global AdS are identical to the QNMs on the AdS Schwarzschild black brane. Motivated by this, in order to examine only the QNMs of large black holes, i.e.~at large temperature, we scale the frequency $\omega$, 
the Wigner-parameter $\mathcal{J}$
and the black hole radius $r_+$ 
as $\alpha$ is taken to $\infty$~\cite{Horowitz:1999jd}:  %
\begin{equation}\label{eq:largeBlackHoleLimit}
    \begin{aligned}
        \omega  &\rightarrow \alpha \left(\frac{r_+}{L}\right) (2 \nu) \, , \\
        \mathcal{J} & \rightarrow \alpha \left(\frac{r_+}{L}\right) j \, , \qquad\qquad \alpha\to \infty\\
        r_+ & \rightarrow \alpha\, r_+ \, ,
    \end{aligned}
\end{equation}
and all other coordinates are not scaled. This  leads us to case \hyperlink{caseii}{case (ii)}. 
Similar to~\cite{Horowitz:1999jd} we argue that $\mathcal{J}$ and $\omega$ must scale like $\alpha$, while $t$ and $r$ scale like $1/\alpha$, so that $D^\mathcal{J}_{\mathcal{KM}}(\theta,\phi,\psi)$ and $e^{-i \omega t}$ do not scale. Finally, take the limit $\alpha \rightarrow \infty$, and keep the leading non-vanishing term in the perturbation equations of motion. 

Now, $j$ can be considered continuous (whereas $\mathcal{J}$ remains discrete). This can be seen  taking the difference of two consecutive values of $j$ before taking the limit: $\Delta j = j_{n+1}-j_n \propto (\mathcal{J}_{n+1}-\mathcal{J}_{n})/\alpha$. For $\alpha\to \infty$, $\Delta j\to 0$. 
This leaves us with large black hole perturbation equations of motion around black holes, i.e.~in case \hyperlink{caseii}{(ii)} discussed in Sec.~\ref{sec:holographicSetup}. 
In this case, we now compute the $h_{++}$ QNMs at $a=0$ as a function of $j$. Comparing with the $h_{x y}$ QNMs of the planar black brane as a function of the momentum $k$, we find excellent agreement within numerical accuracy.
For a complementary view, these QNMs are shown in the complex frequency plane in Fig.~\ref{fig:qnfsh++incompplane}
\begin{figure}
    \centering
    \includegraphics[width=\textwidth]{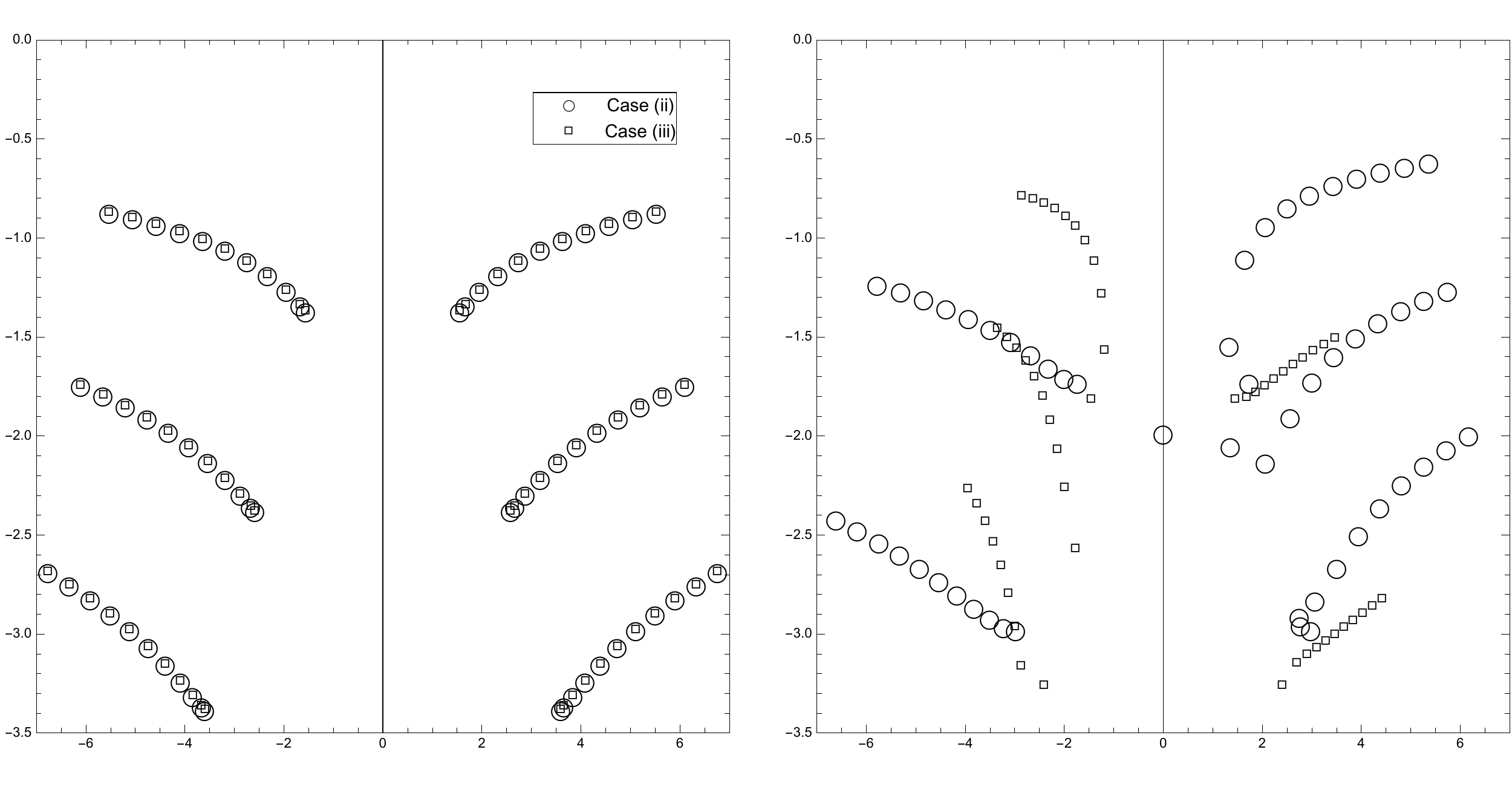}
    \caption{\label{fig:hpphxycompare} {\it Comparison of the QNMs of tensor perturbations $h_{x y}$ vs.~$h_{++}$.} The QNMs of $h_{xy}$ 
    (case~\protect\hyperlink{caseiii}{(iii)}) are shown as squares, those of $h_{++}$ (case~\protect\hyperlink{caseii}{(ii)}) as circles. At $a=0$ (left), the QNMs are identical. For $a\neq0$ (right), the QNMs are not equal.
    }
\end{figure}

\begin{figure}
    \centering
    \includegraphics[width=1\textwidth]{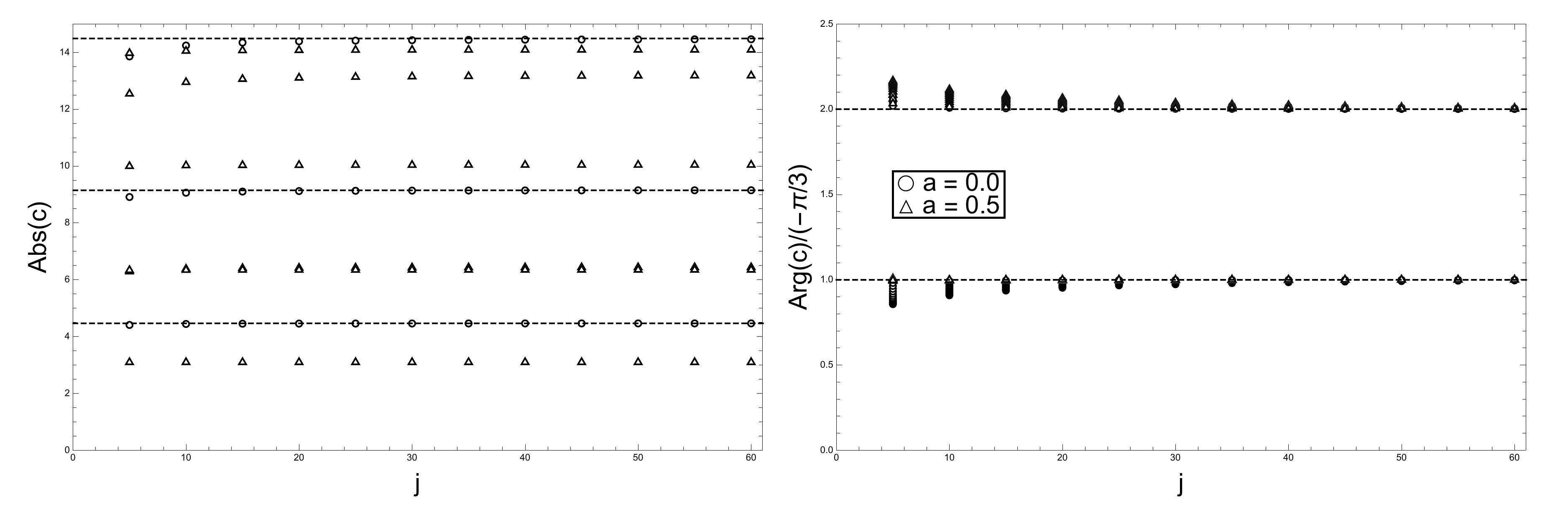}
    \caption{\label{fig:hpplargej}
    {\it Large $j$ (eikonal limit) analysis.} Shown here are the quantities $\text{Abs}(c)$ and $\text{Arg}(c)/(-\pi/3)$, , where $c := \left( \left( \frac{\nu}{j} \right)  - \text{sign}\left(\text{Re}\left( \frac{\nu}{j} \right)\right)\right) \left( 2 j \right)^{4/3} $, derived from the QNMs of $h_{++}$ after the limit~\eref{eq:largeBlackHoleLimit}. The dashed lines are the expected values according to~\cite{Festuccia:2008zx,Fuini:2016qsc}. For large $j$, the values at $a=0$ (circles) and $a=0.5$ (triangles) in the right plot converge to the predicted ones. In the left plot, the same is true for the case $a=0$ (circles). At $a\neq 0$ (triangles), however, we discover a new feature. The large $j$ values shift depending on $a$. }
\end{figure}

In order to further test the role of the angular momentum parameter $j$ of the perturbations, we consider the QNMs from small to very large values of $j$. Under the assumption that $j$ is equivalent to the linear spatial momentum $k$ for the behavior of QNMs, $j\to\infty$ is called the eikonal limit, for which a general analytic expectation for the behavior of QNM frequencies was derived~\cite{Festuccia:2008zx,Fuini:2016qsc}.\footnote{Details of this method can also be found in~\cite{Garbiso:2019uqb}, where we applied it to QNMs of Ho\v rava gravity.} 
The result is shown in Fig.~\ref{fig:hpplargej}. Under the assumption that $j$ is equivalent to the linear momentum $k$, the quantity $c(k)\to c(j)$,  defined in~\cite{Festuccia:2008zx,Fuini:2016qsc}, satisfies the predicted behavior at large values of $j$. 
Our numerical results imply the following large-$j$ expansion for the $n^{\text{th}}$ mode: $2 \nu = 2j+c_n e^{-i (\pi /3)} (2 j)^{-1/3}+O((2 j)^{-1})$, where the conjugate frequencies $-2 \nu^* = -2 j+c_n e^{-i (2\pi /3)} (2 j)^{-1/3}$ are also included in our analysis. The $c_n$ are real numbers in agreement with~\cite{Fuini:2016qsc}. To relate to~\cite{Fuini:2016qsc}, the following identifications are needed: $\omega_F\equiv2\nu$ and $q_F\equiv2 j$. Subscripts ``F'' indicate the Fuini et al. quantities~\cite{Fuini:2016qsc}. 
For large $j$, the values at $a=0$ (circles) and $a=0.5$ (triangles) in the right plot converge to the predicted ones. In the left plot, however, the values of $\text{Abs}(c)$ for large $j$ approach shifted values if $a\neq 0$. 
These observations are the basis for the following hypothesis: 

\begin{center}
    \parbox{0.8\textwidth}{
        In the large black hole  limit~\eqref{eq:largeBlackHoleLimit}, the parameter $j$, measuring the angular momentum of the perturbation, is the Wigner representation analog of the Fourier representation momentum $k$. The perturbation $h_{++}$ is the simply spinning black hole analog of the $h_{xy}$ perturbation around a black brane. This analogy can be extended to all other perturbations.
    }\hypertarget{hyp1} {\hspace{1 cm}} (H.I)
\end{center}

\bigskip 

\noindent In the following sections, we test this hypothesis. For example, in Sec.~\ref{sec:resultsII}, we analytically derive that under the large black hole limit~\eqref{eq:largeBlackHoleLimit}, the $h_{++}$ equation turns into the equation of a minimally coupled massless scalar, which agrees at $a=0$ with the $h_{x y}$ equation on the Schwarzschild black brane~\cite{Policastro:2002se}. 
There should be a direct relation between the eigenfunctions which we consider, namely the Wigner-D functions on $S^3$ and the Fourier modes in the locally flat large black hole limit. Regrettably, we have to come back to this point after a future epiphany.

\subsubsection{Momentum diffusion}
\label{sec:momentumDiff}
\begin{figure}
    \centering
    \includegraphics[width=1\textwidth]{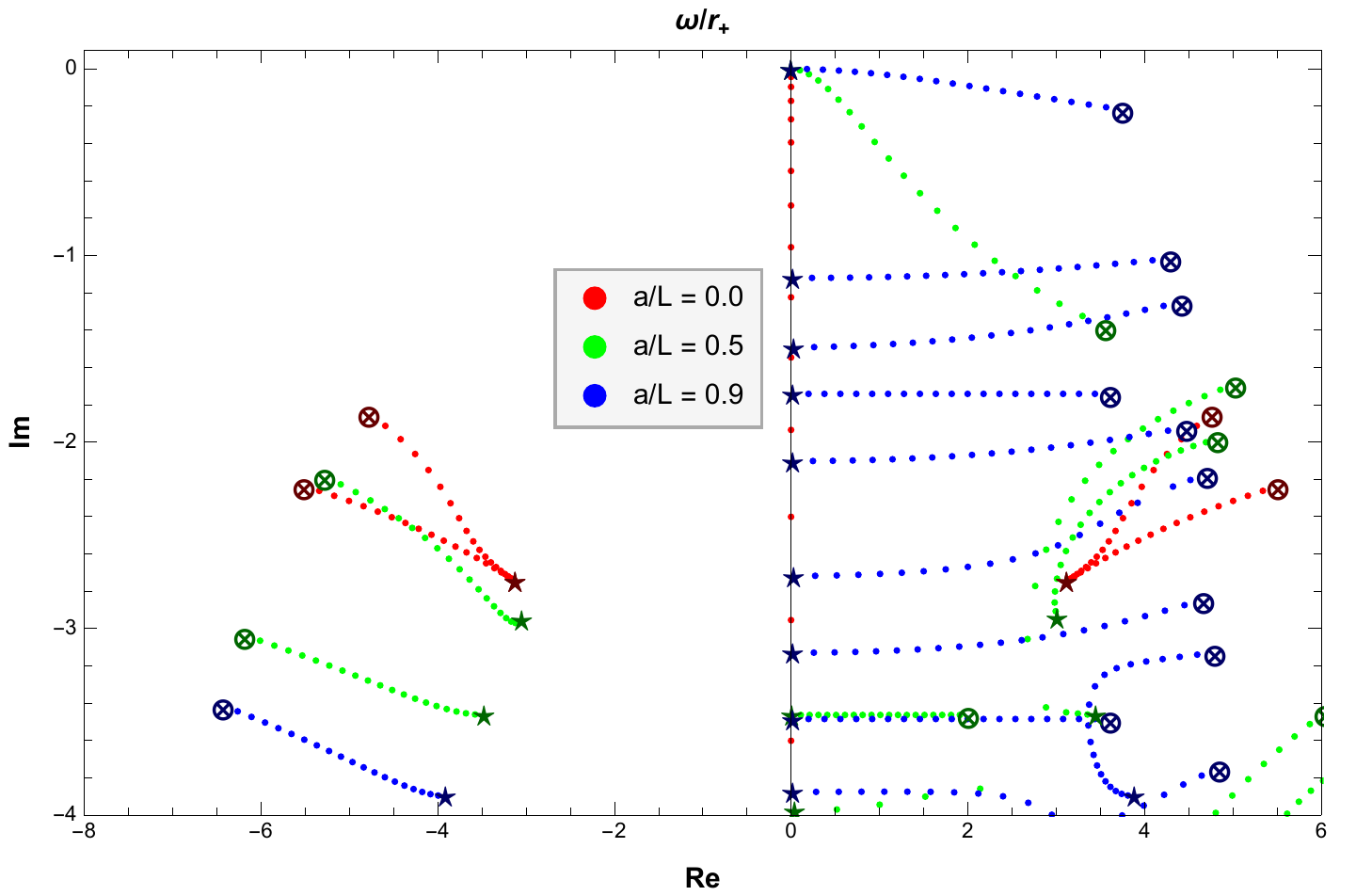}
    \caption{\label{fig:vectqnmsJ}{\it Vector sector QNMs.} The QNMs of $h_{3+}$ and $h_{t+}$ (and $h_{++}$) at $\mathcal{K}=\mathcal{J}+1$ for $r_+ = 100$  at $\mathcal{J}={0,5,10,20,\dots,100}$, where the star indicates the value at $\mathcal{J}=0$, the circled cross indicates $\mathcal{J}=100$. This is case~(i). }
\end{figure}

In analogy to Fig.~\ref{fig:hppQNMsofJ}, in Fig.~\ref{fig:vectqnmsJ}, we show the QNMs of the vector fluctuations $h_{3+},\, h_{t+}$ in the complex frequency plane as a function of the Wigner function parameter $\mathcal{J}$. As a reminder, this is case \hyperlink{caseii}{(i)} defined in Sec.~\ref{sec:holographicSetup}.
It is obvious that there is one mode which satisfies the condition $\omega(\mathcal{J}) \to 0$, as $\mathcal{J}$ is decreased in integer steps from 100 down to 0. 
This condition resembles the standard condition for hydrodynamic modes, $\lim\limits_{k\to 0}\omega(k) = 0$, with spatial momentum $k$. Hence, we refer to this mode as a {\it hydrodynamic mode}. 

Our goal is to analyze this mode in large black holes.
Again, in order to examine only the QNMs of large black holes, we scale the frequency $\omega$, 
the Wigner-parameter $\mathcal{J}$ 
and the black hole radius $r_+$ 
as $\alpha$ is taken to $\infty$, according to Eq.~\eqref{eq:largeBlackHoleLimit}, which leaves us in case \hyperlink{caseii}{(ii)} again.  
Now we define a mode as a hydrodynamic mode in terms of $j$, if it satisfies the condition 
\begin{equation}\label{eq:definitionRotatingHydroMode}
    \lim\limits_{j\to 0}\nu(j) = 0 \, .
\end{equation}

\begin{figure}
    \centering
    \includegraphics[width=1\textwidth]{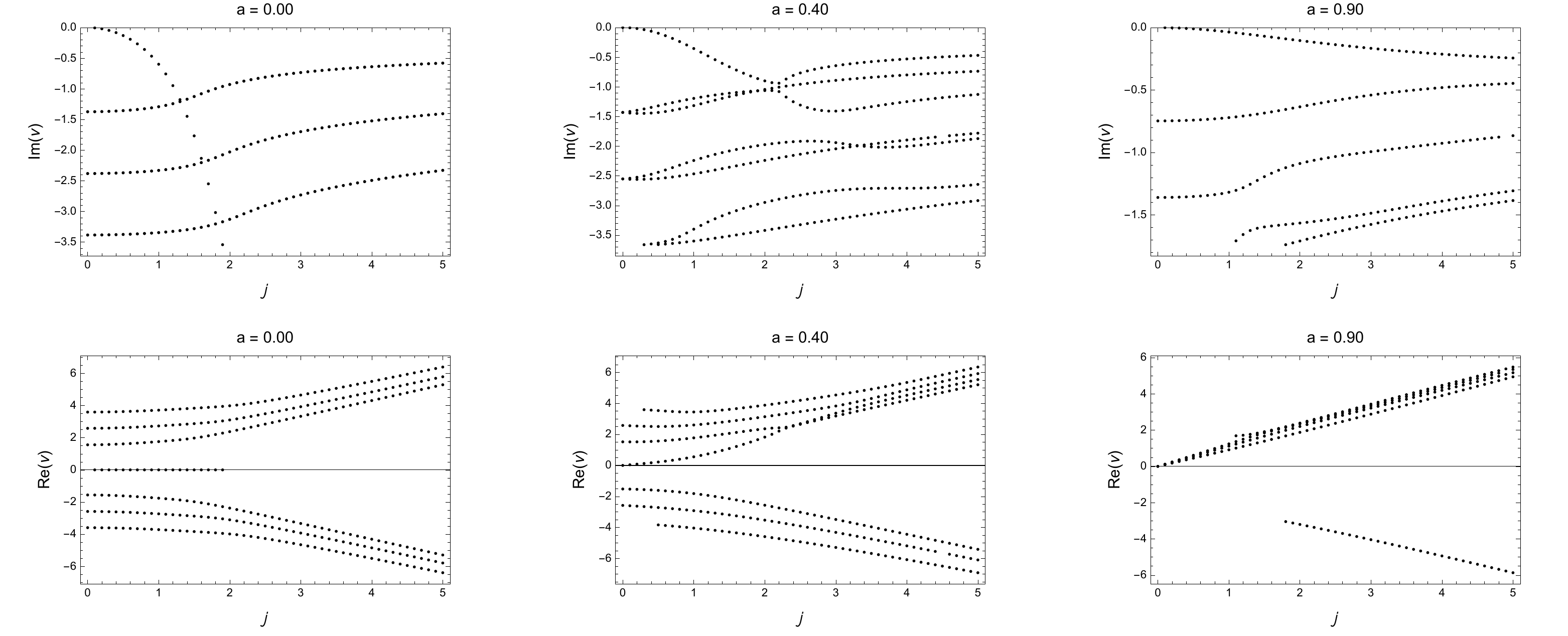}
    \caption{\label{fig:qnfsh3+t+} 
    {\it Vector sector dispersion relations.} Displayed here are the real and imaginary parts of quasinormal frequencies of $h_{3+}$ and $h_{t+}$  for large black holes in case (ii) as a function of $j$.
    Modes displayed here were found with precision of at least $d\nu = \nu 10^{-5}$. Missing modes did not converge to sufficient precision and have been filtered out by the numerical routine.
    For a complementary view, these QNMs are shown in the complex frequency plane in Fig.~\ref{fig:qnfsh3+t+incompplane}.
    }
\end{figure}

Fig.~\ref{fig:qnfsh3+t+} shows the real and imaginary part of this hydrodynamic mode. At small $j\ll 1$, the imaginary part displays a quadratic dependence on $j$ for all values of $a\in [0,1]$. The real part of this hydrodynamic mode shows qualitatively distinct behavior depending on $a$. At $a=0$ it has $\text{Re}\,\nu =0$, but at $a\neq 0$ it has $\text{Re}\,\nu \neq 0$. 

\begin{figure}
    \centering
    \includegraphics[width=0.75\textwidth]{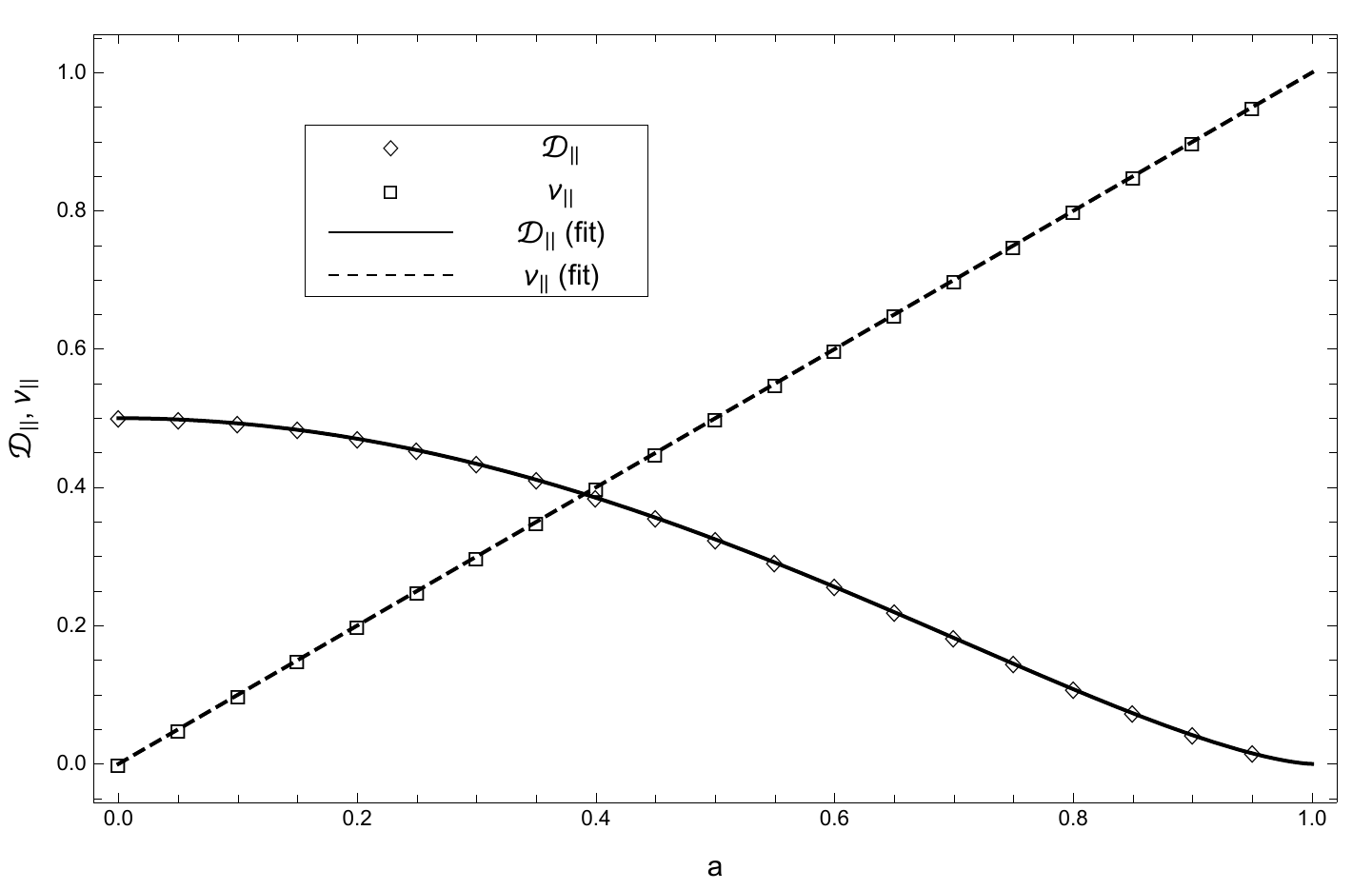}
    \caption{\label{fig:momentumPropagation}
    {\it Momentum diffusion  and propagation in a simply spinning black hole} (case \protect\hyperlink{caseii}{(ii)}).
    The diffusion/damping coefficient, $\mathcal{D}$, and propagation speed, $\mathit{v}$, defined as expansion coefficients of the hydrodynamic vector mode, $\nu \approx \mathit{v}_{||}(a) j-i\mathcal{D}_{||}(a) j^2$. With a $\chi^2\approx10^{-100}$, our fit suggests $\mathit{v}_{||,\text{fit}}=a$ and $\mathcal{D}_{||,\text{fit}}(a)=\frac{1}{2}(1-a^2)^{3/2}$ with $\chi^2 \approx 10^{-100}$.} 
\end{figure}

The quadratic behavior of $\text{Im}\,\nu$ is reminiscent of the momentum diffusion mode in the vector channel of black branes, where the QNM frequency obeys $\omega = -i k^2 \mathcal{D} +\mathcal{O}(k^4)$~\cite{Policastro:2002se,Kovtun:2005ev}. Hence, we refer to this QNM in $(h_{t+},h_{3+})$ as {\it momentum diffusion mode}.  With this in mind, we fit our data to the ansatz $\nu = v_{||} j - \mathcal{D}_{||} j^2$, in the range $j=0, ..., 0.1$. 
Then, $\mathcal{D}_{||}$ is of order 1 and is displayed as a function of the angular momentum parameter $a$ in Fig.~\ref{fig:momentumPropagation}. Note that at $a=0$ the black brane value $\mathcal{D}_{\text{PSS}}=\mathcal{D}_{||}(a=0)/(2\pi T)=1/(4\pi T_0)$ is recovered ($r_+=1$ was chosen, such that $T=1/\pi$), where the subscript ``0'' indicates $a=0$.\footnote{
The subscript ``PSS'' refers to the quantities defined in~\cite{Policastro:2002se}. Note that the $\omega_\text{PSS} = 2 \pi T \nu L$ and $q_\text{PSS} = 2 \pi T j L$.} In that case, it is well-known that the momentum diffusion coefficient is related to the shear viscosity by $\mathcal{D}_{\text{PSS}}=\mathcal{D}_{||}(a=0)/(2\pi T)=\eta_0/(\epsilon_0 + P_0)$. Below, in Sec.~\ref{sec:resultsII}, we will generalize this relation to the rotating case $a\neq 0$.

Remarkably, this momentum diffusion mode turns into a propagating mode at $a\neq 0$. Our fit yields a nonzero $v_{||}$, which we interpret as the speed with which this mode propagates. Fitting our numerical results, we conjecture the momentum diffusion mode turns into a propagating mode with velocity and damping coefficient analytically given by
\begin{eqnarray}\label{eq:diffusionMode}
    \nu &=& {v}_{||} j - \mathcal{D}_{||} j^2 \, , \\
    v_{||} &=& a \, , \\
    \mathcal{D}_{||} &=&  (2\pi T)\, \mathcal{D}_{\text{PSS}} \, (1-a^2)^{3/2} = \frac{1}{2} (1-a^2)^{3/2} \, .
\end{eqnarray}
Note that this mode can be viewed as a non-propagating diffusion mode when defining the frequency $\bar\nu=\nu- j a$, which is also discussed as the natural choice of frequency in~\cite{Ishii:2018oms}.\footnote{If we consider $a$ as a ``chemical potential for rotation'', this shift is reminiscent of the frequency shift due to an isospin chemical potential~\cite{Erdmenger:2007ja,Kaminski:2008ai}.}
Then, we obtain the  diffusion mode as a function of the angular momentum parameter $a$ with
\begin{eqnarray}\label{eq:diffusionModeBarNu}
    \bar\nu &=& \bar{v}_{||} j - \mathcal{D}_{||} j^2 \, , \\
    \bar{v}_{||} &=& 0 \, , \\
    \mathcal{D}_{||} &=& \frac{1}{2} (1-a^2)^{3/2} \, .
\end{eqnarray}
In~\cite{Ishii:2018oms}, the definition of $\bar\nu$ is part of a coordinate system in which the AdS-boundary is not rotating. In our coordinates, our background is already non-rotating at the AdS-boundary. Instead, here the perturbations appear to be rotating at the AdS-boundary. Hence, we adopt this definition of $\bar\nu$ simply in order to define a frequency and time coordinate in which the momentum diffusion mode is non-propagating.

The dispersion relation given in \eref{eq:diffusionMode} can be related to that of a fluid at rest. Such a relation was found for a fluid moving with {\it linear} velocity $\mathbf{v}_0$~\cite{Kovtun:2019hdm,Hoult:2020eho}\footnote{See also~\cite{Abbasi:2017tea}.}
\begin{equation}\label{eq:movingFluidDiffusion}
    \omega = \mathbf{v}_0\cdot \mathbf{k}- i D \sqrt{1-\mathbf{v}_0^2} (\mathbf{k}^2-(\mathbf{v}_0\cdot \mathbf{k})^2)+\mathcal{O}(\mathbf{k}^3)\, , 
\end{equation}
if the fluid at rest has a diffusion mode $\omega = - i D \mathbf{k}^2+\mathcal{O}(\mathbf{k}^3)$. 
Obviously, this dispersion relation depends on the angle, $\theta$, between the {\it linear} velocity $\mathbf{v}_0$ and the momentum, $\mathbf{k}$, through the scalar product, $\mathbf{v}_0\cdot \mathbf{k} = |\mathbf{v}_0|\,|\mathbf{k}| \cos{\theta}$. 
Inspired by this relation, we generalize the relation between transport coefficients in rotating fluids and fluids at rest under two assumptions.  
First, we assume that~\eref{eq:movingFluidDiffusion} can be generalized from a linear fluid velocity, $\mathbf{v}_0$, to an angular fluid velocity $\mathbf{w}$. Then, identify $a$ as the only nonzero element of that angular fluid velocity, replacing $\mathbf{v}_0\to \mathbf{w}=(0,0,v_0=a)$. Second, we assume $\mathbf{w}$ is aligned with $\mathbf{k}\to \mathbf{j} = (0,0, 2\pi T \,j)$.\footnote{Like the original Wigner parameter $\mathcal{J}$, also $j$ is describing a perturbation with angular momentum in the 3-direction. 
 } 
Finally, $\omega\to 2\pi T \, \nu$. 
Then indeed our dispersion relation \eref{eq:diffusionMode} is given by \eref{eq:movingFluidDiffusion}. 

This leads us to conjecture a general formula for the diffusion dispersion relations for arbitrary angle, $\theta$, between the angular velocity three-vector, $\mathbf{w}$, and the mode's angular momentum three-vector, $\mathbf{j}$:
\begin{equation}
    \label{eq:rotatingFluidDiffusion}
    \omega = \mathbf{w}\cdot (2\pi T\mathbf{j})- i D \sqrt{1-\mathbf{w}^2} \left((2\pi T\mathbf{j})^2-(\mathbf{w}\cdot (2\pi T\,\mathbf{j}))^2\right)
    +\mathcal{O}(\mathbf{j}^3)\, ,
\end{equation}
with the diffusion coefficient $D$ taking its value at rest. 
This relation yields the correct result for the limiting case, Eq.~\eqref{eq:diffusionMode} for $\theta=0$, and presumably also for the case $\theta=\pi/2$ when comparing to the linear velocity results in~\cite{Kovtun:2019hdm,Hoult:2020eho}.  
Testing this relation for general $0<\theta \le \pi/2$ would require breaking all remaining symmetries. In that technically challenging case all the excitations couple to each other, requiring a separate analysis. 
This is beyond the scope of this work and presents a direction for future investigations.

Notably, in our case, the fluid is rotating not translating. 
We stress that this makes our results in cases \hyperlink{caseii}{(i)} and \hyperlink{caseii}{(ii)} different from a boosted black brane, case \hyperlink{caseii}{(iii)}. As an example, the diffusion coefficient for a boosted black brane is different from that of our large rotating black hole, see Fig.~\ref{fig:diffusionCoefficient}.

\subsubsection{Sound propagation}
\begin{figure}
    \centering
    \includegraphics[width=1\textwidth]{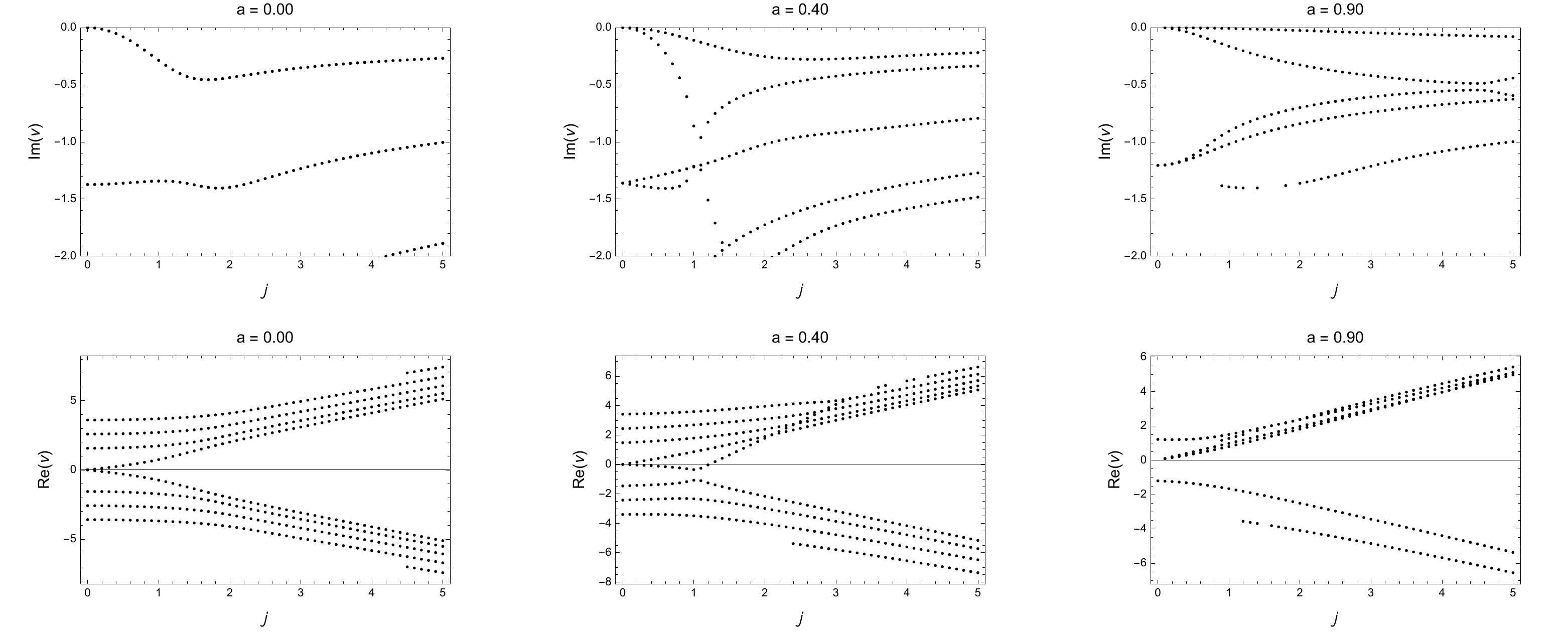}
    \caption{
    {\it Scalar QNMs of the simply spinning black hole.}
    There are two hydrodynamic modes, which we identify as sounds modes, which mirror each other regarding their real part at $a=0$, and both have the same imaginary part. At $a=0$ these two modes behave differently. 
    Modes displayed here were found with precision of at least $d\nu = \nu 10^{-5}$. Missing modes did not converge to sufficient precision and have been filtered out by the numerical routine.
    } 
    \label{fig:qnfsscalar}
\end{figure}

\begin{figure}
    \centering
    \includegraphics[width=\textwidth]{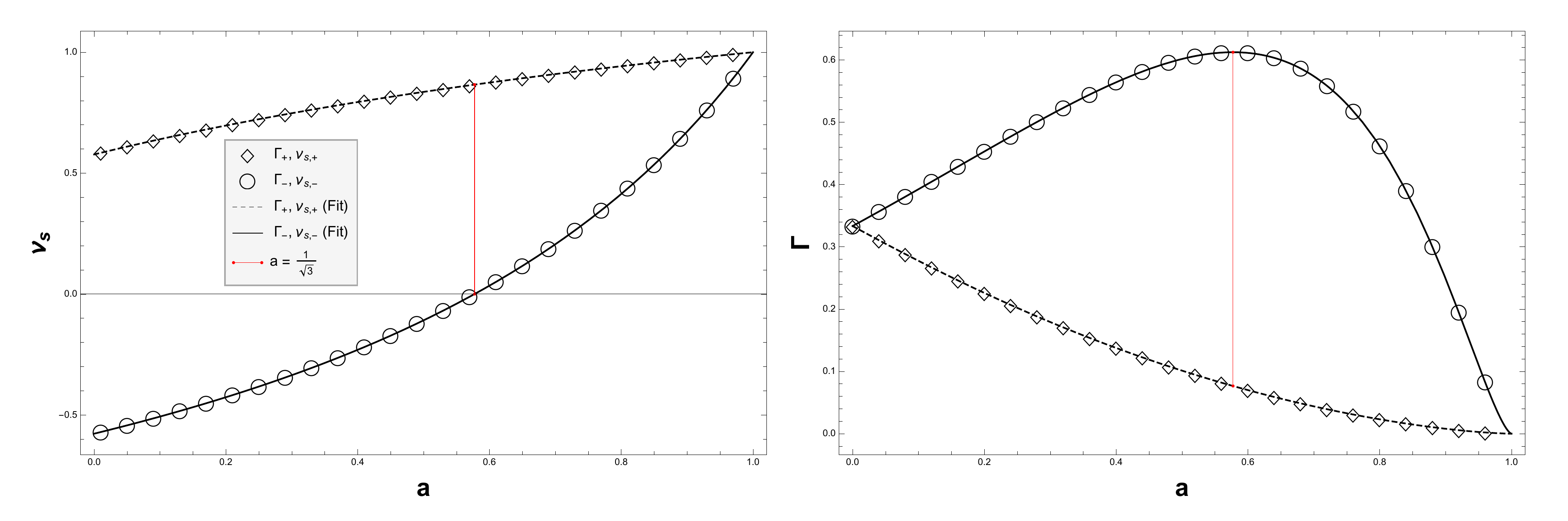}
    \caption{
    \label{fig:soundPropagation}
    {\it Sound propagation in a simply spinning black hole.} Speed of sound, $v_s$, and sound attenuation, $\Gamma$, shown as a function of the angular momentum parameter $a$ of the rotating black hole.
    There are two speeds of sound, $v_{s,\pm}$, and attenuations, $\Gamma_{\pm}$, which are defined through the small $j$ expansion  $\nu(j)=v_{s,\pm}(a)  j - \Gamma_\pm(a) j^2+\mathcal{O}(j^3)$. The fits found
    show that $v_{s,\pm, \text{fit}} = \frac{a\pm\frac{1}{\sqrt{3}}}{1\pm\frac{a}{\sqrt{3}}}$ with $\chi^2 \approx 10^{-100}$ and $\Gamma_{s,\pm, \text{fit}} = \frac{1}{3}\frac{\left(1-a^2\right)^{3/2}}{\left(\frac{a}{\sqrt{3}}\pm1\right)^3}$ with $\chi^2 \approx 10^{-100}$. 
    }
\end{figure}

In the scalar perturbation sector, we find two hydrodynamic modes, as shown in Fig.~\ref{fig:qnfsscalar}. In analogy to the scalar sector of black branes~\cite{Kovtun:2005ev,Policastro:2002tn}, we call them {\it sound modes}. At vanishing rotation, $a=0$, and in the regime $j\le 1$, the real part of both QNMs increases approximately linearly with $j$, while the imaginary part increases approximately quadratically with $j$. 
Motivated by these observations, we fit these two scalar hydrodynamic modes to the ansatz 
$\nu = v_s j - i \Gamma k^2$ in the range $j=0, ..., 0.1$. 
At $a=0$, we find a real-valued $v_s \approx \pm 1/\sqrt{3}$, 
as expected for a speed of sound in a conformal field theory. We also obtain $\Gamma\neq 0$ with $\Gamma\approx 1/3$.
Moving to nonzero angular momentum, $a>0$, we observe that the two speeds of sound with initial values $\pm 1/\sqrt{3}$ change differently with $a$, as seen in Fig.~\ref{fig:soundPropagation}. With increasing angular momentum, both speeds increase. Around $a\approx 1/\sqrt{3}$, this causes the formerly negative speed to vanish and then grow further. Near extremality $a\to 1$, both speeds are asymptotic to the speed of light~$v_{s,\pm}\to 1$. 

Similarly, in Fig.~\ref{fig:soundPropagation}  there are two distinct sound attenuation coefficients, $\Gamma_\pm$. At $a=0$, they start with the same value, at $a>0$ one increases, the other decreases. While $\Gamma_+$ assumes a maximum around $a\approx 1/\sqrt{3}$, $\Gamma_-$ has no extremum. Near extremality, $a\to 1$, both approach zero, $\Gamma_\pm \to 0$. However, while $\Gamma_-$ smoothly asymptotes to zero, i.e.~its slope is going to zero, $\Gamma_+$ approaches zero with a non-vanishing slope. The same behavior is observed in $v_{s,\pm}$ as they approach extremality. 

Similar to the momentum diffusion dispersion above, we also find analytic expressions for the sound modes here fitting to our numerical data
\begin{eqnarray}
\label{eq:soundDispersion}
 \nu &=& v_{s,\pm} \, j - i \Gamma_{s,\pm} \,j^2  \, , \\
 v_{s,\pm} & = & \frac{a\pm\frac{1}{\sqrt{3}}}{1\pm \frac{a}{\sqrt{3}}}\, ,\\
 \Gamma_{s,\pm} & = & \frac{1}{3}\frac{\left(1-a^2\right)^{3/2}}{\left(1\pm \frac{a}{\sqrt{3}}\right)^3} \, ,
\end{eqnarray}
such that $\Gamma_0:=\Gamma_s|_{a=0}=1/3$. 
Under the same two assumptions we stated for the momentum diffusion dispersion relation above, we now compare \eref{eq:soundDispersion} with the sound dispersion relation for a moving fluid~\cite{Kovtun:2019hdm,Hoult:2020eho}, which is given by
\begin{equation}\label{eq:movingFluidSound}
    \omega = \frac{v_0\pm v_s}{1\pm v_0 v_s} k - \frac{i}{2} {\Gamma_0} \frac{(1-v_0^2)^{3/2}}{(1\pm v_0 v_s)^3} k^2 + \mathcal{O}(k^3) \, , 
    \quad \text{for}\, \, \mathbf{k}||\mathbf{v}_0 \, ,
\end{equation} 
if the fluid at rest has two sound modes $\omega = \pm v_s k - i k^2 \Gamma_0 +\mathcal{O}({k}^3)$. Here we have taken into account the distinct notation  $\Gamma_0=\gamma_s/(2(\epsilon_0+P_0))$ because in~\cite{Kovtun:2019hdm} $\gamma_s$ is used. 
Now identify $v_0\to a$, $\omega\to 2\pi T \, \nu$ and $k\to 2\pi T \, j$ and we obtain \eref{eq:soundDispersion}. 

Thus, also for the sound dispersion relation in a rotating fluid we conjecture two more general equations. If the angular velocity of the fluid, $\mathbf{w}$, and the angular momentum of the mode, $\mathbf{j}$ are aligned, we conjecture
\begin{equation}\label{eq:rotatingFluidSoundParallel}
     \omega = \frac{a\pm v_s}{1\pm a v_s} (2\pi T\, j) - \frac{i}{2} {\Gamma_0} \frac{(1-a^2)^{3/2}}{(1\pm a v_s)^3} (2\pi T\, j)^2 + \mathcal{O}(j^3) \, , 
    \quad \text{for}\, \, \mathbf{j}||\mathbf{w} \, ,
\end{equation}
with the sound speed, $v_s$, and the sound attenuation, $\Gamma_0$, taking their values at rest. 
Again, we generalize this equation to the perpendicular case between $\mathbf{w}$ and $\mathbf{j}$ in analogy to the corresponding equations for {\it linear} fluid motion derived in~\cite{Hoult:2020eho,Kovtun:2019hdm}. So we conjecture:
\begin{equation}\label{eq:rotatingFluidSoundPerpendicular}
     \omega = \pm \frac{(1-a^2)^{1/2}}{(1- a^2 {v_s}^2)^{1/2}} v_s \, 2\pi T\,j 
     - \frac{i}{2} {2 \Gamma_0} \frac{(1-a^2)^{1/2}}{(1- a^2 v_s^2)^2} (2\pi T \, j)^2 + \mathcal{O}(j^3) \, , 
    \quad \text{for}\, \, \mathbf{j}\perp\mathbf{w} \, .
\end{equation}
We are tempted to generalize even further to arbitrary angles. However, already the angle dependent expressions for {\it linear} fluid motion in~\cite{Hoult:2020eho,Kovtun:2019hdm} are involved and an adaption to the {\it rotating} case is not obvious to us without technically involved extensions of our work.

In the case $a=0$ it is well-know that $\Gamma$ is related to $\eta$ by~\cite{Policastro:2002tn}\footnote{There is an overall factor of 2 compared to the conventions of Policastro, Son, and Starinets~(PSS)~\cite{Policastro:2002tn}, because our dimensionless $j=k/(2\pi T)$, such that our $\Gamma_0=2 \Gamma_{
\text{PSS}}$.}
\begin{equation}
\label{eq:soundAttenuation0}
    \Gamma_0 = 2 \frac{\zeta_0+4\eta_0/3}{2(\epsilon_0 + P_0)} \overset{\text{\tiny conformal}}{=} 2 \frac{2\eta_0}{3(\epsilon_0 + P_0)} \, ,
\end{equation}
where the bulk viscosity $\zeta=0$ in conformal field theories. We will generalize this relation to the rotating case $a\neq 0$ below.

\subsection{QNMs of planar limit black brane}
\label{sec:planarLimit}
\begin{figure}
    \centering
    \includegraphics[width=1\textwidth]{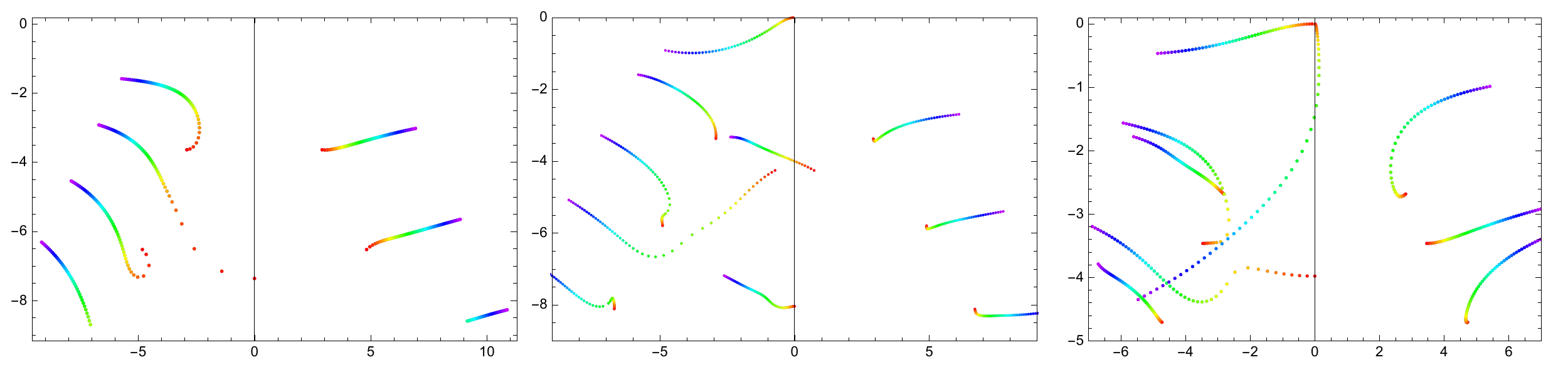}
    \caption{\label{fig:sampleplanarqnfs} 
    {\it Planar limit black brane QNMs (case \protect\hyperlink{caseiii}{(iii)}).} 
    The planar brane QNMs are displayed in the complex frequency plane for $a=0.5$. All the sectors are represented from left to right: tensor, vector, scalar. The linear momentum coers the range $0<k<5$, indicated by the color (red to violet). The units are chosen such that $r_h=L=1$.}
\end{figure}

For comparison, we now consider the planar limit black brane, case \hyperlink{caseiii}{(iii)}.  Fig.~\ref{fig:sampleplanarqnfs} shows examples for the QNMs in all sectors (tensor, vector, scalar). Clearly, the nonzero angular momentum $a$ affects the QNM spectra. Most prominently, there is no mirror symmetry about the vertical (imaginary frequency) axis as was the case at $a=0$. The angular momentum breaks parity and separates modes into those which are rotating with and those rotating against the black hole / fluid. 

\begin{figure}
    \centering
    \includegraphics[width=0.75\textwidth]{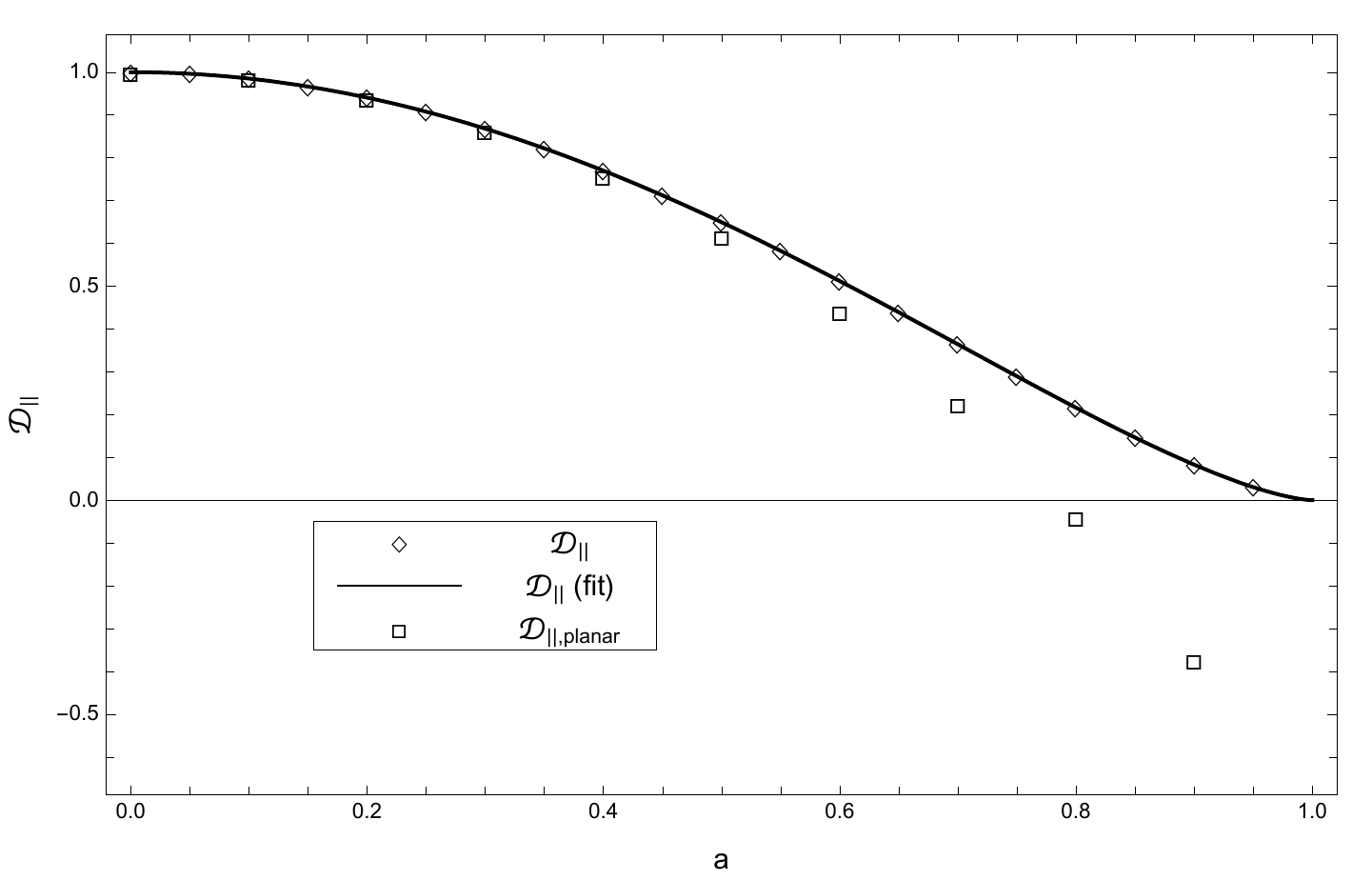}
    \caption{\label{fig:diffusionCoefficient}
    {\it Momentum diffusion in a simply spinning black hole versus the planar limit black brane.} 
    Displayed is the diffusion coefficient $\mathcal{D}_{||}$ extracted in the momentum diffusion (vector perturbation) sector from the hydrodynamic QNM of the field $h_{+3}$ (circles) of the black hole. It is compared to the diffusion coefficient of the field $h_{xy}$ (dots) of the planar limit black brane. The planar limit breaks down around $a=0.7$, where the planar limit black brane QNM acquires a positive imaginary part, rendering that planar limit black brane unstable. The black hole however is stable throughout the whole regime from $a=0$ to (near) extremality $a\approx 1$. 
    }
\end{figure}

As we had seen in Fig.~\ref{fig:hpphxycompare}, the QNMs of the planar limit black brane agree with those of the simply spinning black hole at $a=0$, but differ at $a\neq 0$. This difference shows itself more drastically in the momentum diffusion (vector) channel. The longitudinal momentum diffusion coefficient of the planar black brane (case \hyperlink{caseiii}{(iii)}) is compared to that of the simply spinning black hole (case \hyperlink{caseii}{(ii)}) in Fig.~\ref{fig:diffusionCoefficient}. Around $a\approx 0.8$, the  the momentum diffusion coefficient becomes negative. This indicates that the corresponding QNM acquires a positive imaginary part, leading to exponential growth, i.e.~signaling an instability. The simply spinning black hole, however, remains stable throughout the whole range of $a$ from zero to extremality.

Despite the $h_{++}$ QNMs being different from the $h_{x y}$ QNMs, we will show below that both of the shear viscosities extracted from the planar limit black brane do agree with the spinning black hole values for all $a$. 

\section{Results II: Shear Viscosities}
\label{sec:resultsII}
%
\subsection{Longitudinal shear viscosity of spinning black hole} \label{sec:longituidnalShear}
In this section, we compute the correlators in the vector channel, i.e.~$\langle T_{+3} T_{+3} \rangle$, $\langle T_{+3} T_{+t} \rangle$,  $\langle T_{+t} T_{+3} \rangle$, $\langle T_{+t} T_{+t} \rangle$. For the black brane, this would be the momentum diffusion sector, from which one could extract a shear viscosity and a momentum diffusion constant. Following our hypothesis~\hyperlink{hyp1}{(H.1)}, we now attempt to extract these hydrodynamic quantities from the analogous correlators of the black hole case. 
For this purpose, we follow the standard holographic procedure for computing correlation functions from coupled gravitational fluctuation equations outlined in appendix~\ref{sec:correlators}. Given previous Kubo relations derived or simply used in anisotropic systems~\cite{Rebhan:2011vd,Ammon:2017ded,Ammon:2020} we base our shear viscosity results in this section on a second hypothesis:
\begin{center}
    \parbox{0.8\textwidth}{
        We assume that at $a\neq 0$, but vanishing angular momentum of the perturbation, $j=0$, the Kubo formulae for the two shear viscosities are only mildly modified by the angular momentum anisotropy, and $\eta_\perp$ is related to the tensor correlator $\langle T_{++} T_{++} \rangle$, while $\eta_{||}$ is related to the vector correlator $\langle T_{+3} T_{+3} \rangle$ in analogy to~\cite{Rebhan:2011vd}. 
    }\hypertarget{hyp2} {\hspace{0.6 cm}} (H.II)
\end{center}
We find that at $j=0$ and to linear order in $\nu$ the energy-momentum tensor correlators in this sector behave as follows
\begin{eqnarray}
    \langle T_{+3} T_{+3} \rangle = -i \nu \eta_{||} + \mathcal{O}(\nu^2,j)  \, ,\\ 
    \langle T_{+t} T_{+3} \rangle = 0 + \mathcal{O}(\nu^2,j)  \, ,\\ 
    \langle T_{+3} T_{+t} \rangle = 0 + \mathcal{O}(\nu^2,j) \, ,\\ 
    \langle T_{+t} T_{+t} \rangle =  0 + \mathcal{O}(\nu^2,j)\, .
\end{eqnarray}
This is reminiscent of the correlation functions from~\cite{Policastro:2002se}, $\langle T_{x3} T_{x3} \rangle \propto \omega^2$,  $\langle T_{x3} T_{tx} \rangle \propto \omega k$, 
$\langle T_{tx} T_{tx} \rangle \propto k^2$, in the case $k=0$. Here we have chosen the notation from~\cite{Policastro:2002se}, where $\omega$ and $k$ are the frequency and momentum in $z$-direction (which is equal to $z$ in our notation), associated with the Fourier transform $\propto e^{-i \omega t + i k z}$. Summarizing this point, we find that in the hydrodynamic limit $\nu\ll 1$ 
and at $j=0$ the correlators of the spinning black hole behave as if $j$ was analogous to the spatial momentum $k$ in the planar case. 

Now we test our hypothesis~\hyperlink{hyp1}{(H.1)}, interpreting the parameter $j$ as the Wigner representation analog of the Fourier representation momentum $k$. Applying the standard Kubo formula to our two-point function under this assumption, we obtain what we will, in analogy to~\cite{Policastro:2002se},  suggestively refer to as the {\it longitudinal shear viscosity of the rotating black hole}
\begin{equation}\label{eq:etaParallel}
    \eta_{||} = \lim\limits_{\nu\to 0}\frac{\langle T_{+3} T_{+3} \rangle (\omega, j=0)}{-i \nu} \, .
\end{equation}
In our anisotropic system, this longitudinal shear viscosity $\eta_{||}$ is distinct from the transverse shear viscosity $\eta_\perp$, in analogy to the anisotropic system studied in~\cite{Rebhan:2011vd}. 
We were unable to obtain an analytic solution for this correlator, but fitting the numerical data we conjecture the following closed form
\begin{equation}\label{eq:etaParallelConjectured}
    \eta_{||} = \frac{N^2 \pi T_0^3}{8} \sqrt{1-a^2}
    = \frac{N^2 \pi T(a)^3}{8 (1-a^2)} \, .
\end{equation}
Dividing by the entropy density, we obtain
\begin{equation}
    \frac{\eta_{||}}{s} = \frac{1}{4 \pi} (1-a^2)\, .
\end{equation}
Similar to other cases with anisotropies~\cite{Erdmenger:2010xm,Rebhan:2011vd,Critelli:2014kra,Ammon:2020}, this longitudinal shear viscosity over entropy density ratio does not evaluate to $1/(4\pi)$ for all $a\neq 0$. 
And there is no reason for that. 
It is true that for Einstein gravity in a state which is thermal and symmetric under $O(3)$ rotations in the spatial boundary $(x,y,z)$-directions, we always get $\eta/s = 1/(4\pi)$ in the large $N$ limit, as proven in~\cite{Buchel:2003tz,Kovtun:2003wp,Policastro:2001yc}. However, these universality proofs do not apply to states in which the $O(3)$ rotation symmetry is broken. For example, \cite{Buchel:2003tz} assumes that the energy momentum tensor components satisfy $T_{tt} + T_{xx} = 0$. However, in a system with rotation symmetry broken from $O(3)$ down to $O(2)$, e.g. a rotating black hole, there are two distinct $T_{||}=T_{zz}\neq T_{\perp} = T_{xx}$. Hence, the relation above can only be satisfied for one of the directions, either for $T_{||}$ or for $T_{\perp}$. This is why only one of the shear viscosities has to saturate the bound $\eta_\perp/s = 1/(4\pi)$, while the other, $\eta_{||}$, may violate the bound. 

A complementary point of view is that the equation of motion for a given metric perturbation, e.g.~$h_{xz}$, determines the value of the shear viscosity, e.g.~$\eta_{xzxz}$. If that fluctuation transforms as a tensor under $O(3)$ rotations, then it decouples from all other fluctuations and satisfies the equation of motion of a massless scalar. However, if rotation symmetry in the $(x,z)$-plane is broken, e.g.~by an angular momentum in $z$-direction, then the fluctuation $h_{xz}$ transforms like a vector under the remaining $O(2)$ symmetry, and hence couples to the other vector perturbations. This modifies the solution for $h_{xz}$ and changes the value of $\eta_{xzxz}$ compared to $\eta_{xyxy}$.  

It turns out that we can relate this shear viscosity to the momentum diffusion coefficient computed in Sec.~\ref{sec:momentumDiff} through an angular momentum dependent Einstein relation. 
The diffusion coefficient at $a=0$ is given by~\cite{Policastro:2002se} 
\begin{equation}\label{eq:momentumDiffusionCoefficientNonrotating}
    \mathcal{D}_{\text{PSS}}=\frac{\eta_0}{\epsilon_0 + P_0}= \frac{\eta_0}{s_0} \frac{1}{T_0}=\frac{1}{4\pi}\frac{1}{T_0} \, ,
\end{equation} 
where the subscripts ``0'' indicate that these quantities are evaluated in a non-rotating state, $a=0$. Here we have used $\epsilon_0+P_0 = s_0 T_0$. At nonzero $a$, however, the latter relation as well as $\eta$ are modified.  
\begin{equation}
\label{eq:DLFit}
    \mathcal{D}_{||}(a)
    = 2 \pi T_0 \frac{\eta_0}{\epsilon_0 + P_0} \left (
    1-a^2\right)^{3/2}
    \, ,
\end{equation}
In order to generalize  Eq.~\eqref{eq:momentumDiffusionCoefficientNonrotating} to nonzero angular momentum, 
we note that numerically $\epsilon+P_\perp=\frac{s_0 T_0 L}{1-a^2/L^2}$. Hence, we conjecture the following analytic generalization 
\begin{equation}
\label{eq:DL}
    \mathcal{D}_{||}(a)
    =2 \pi T_0 \frac{\eta_{||}(a)}{\epsilon(a) + P_\perp(a) 
    }
    = 2 \pi T_0 \frac{\eta_0}{\epsilon_0 + P_0} \left (
    1-a^2\right)^{3/2} 
    \, ,
\end{equation}
which is reminiscent of the relation found for an anisotropic plasma in~\cite{Rebhan:2011vd}.

We can also cross-check our shear viscosities with the sound attenuation parameter $\Gamma$, extracted from the sound propagation sector. Based again on our numerical data, we conjecture the following generalization of the well-known relation~\eqref{eq:soundAttenuation0} to the relation between sound attenuation and shear viscosity in a system with angular momentum parameter $a$

\begin{equation}
\label{eq:soundAttenuation}
    \Gamma_\pm(a)=\frac{2\eta_{||}(a)}{3(\epsilon(a) + P_\perp(a))} \frac{1}{(1\pm a/\sqrt{3})^3} \, ,
\end{equation}
where $\zeta(a)=0$ for our conformal field theory and we can not include it in this relation. 

\begin{figure}
    \centering
    \includegraphics[width=1\textwidth]{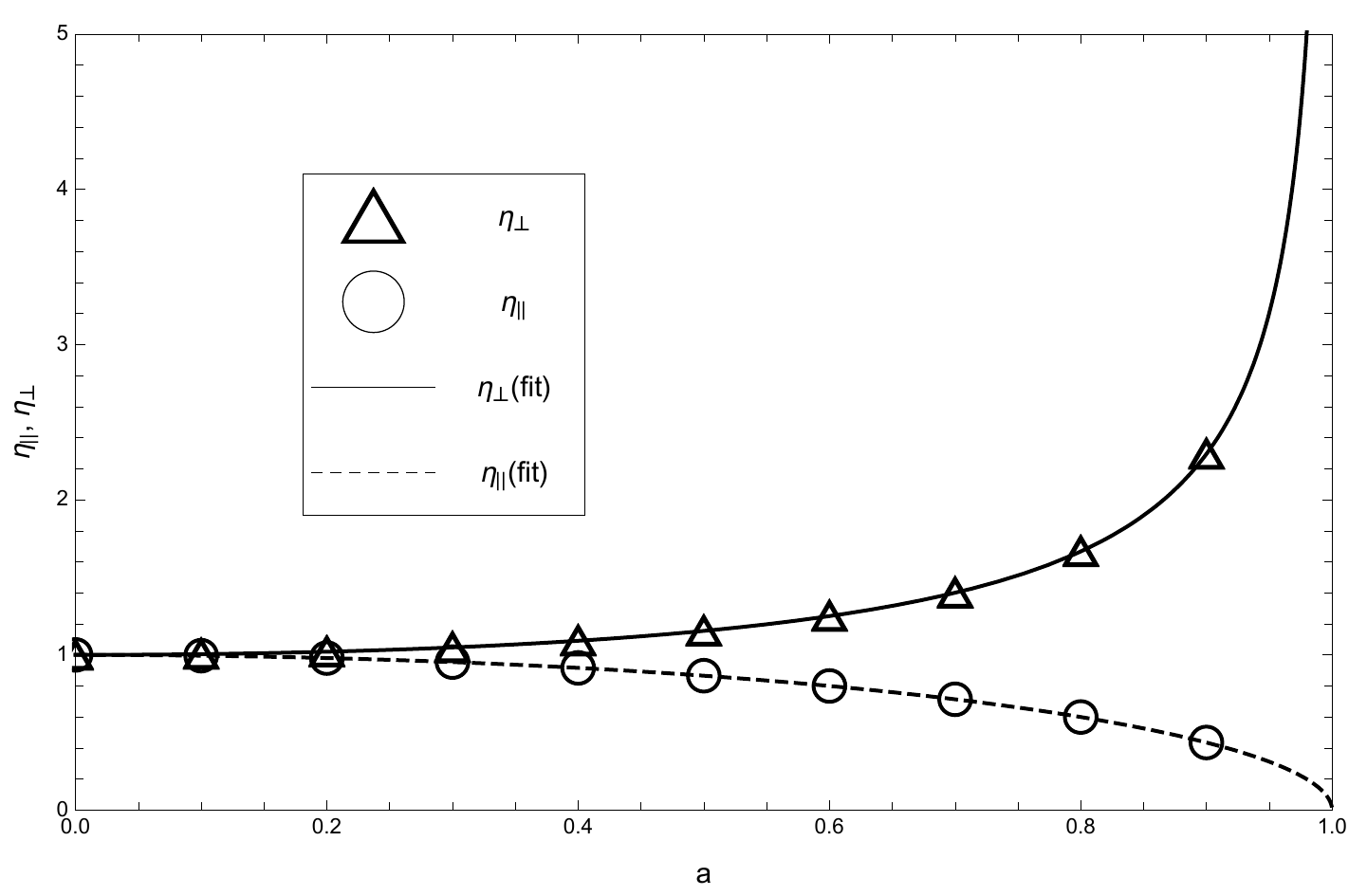}
    \caption{
    {\it Longitudinal and transverse shear viscosities of the simply spinning black hole.}
    $\eta_{\perp}$ and $\eta_{||}$ extracted from numerics are shown as a function of the normalized angular momentum, $a$, of the fluid / black hole. The shear viscosities of the spinning planar limit black brane are coindicent with those shown (shown analytically for $\eta_\perp$, numerically shown for $\eta_{||}$).
    }
\end{figure}


\subsection{Transverse shear viscosity of spinning black hole}

Later in this section, we will provide values for the quasinormal modes of tensor perturbations of the simply spinning black hole. These values will be computed numerically because in general the fluctuation equations are too complicated for an analytic treatment. Before going down that dark and perilous path losing analytic control, however, we begin here deriving some exact results.

\paragraph{Analytic results} The equation of motion for $h_{++}$ decouples from all others if the Wigner-parameters satisfy the relation $\mathcal{K}=\mathcal{J}+2$. This is reminiscent of the tensor (helicity-2 under spatial $O(3)$-rotations) fluctuation in the planar case, which is also referred to as ``scalar'', because it satisfies the equation of a massless scalar in the respective spacetime, e.g.~AdS Schwarzschild. 

Let us begin our considerations at vanishing angular momentum, $a=0$, i.e.~in global Schwarzschild $AdS_5$. After a Fourier transform of the time-coordinate the equation of motion for $h_{++}$ given in Eq.~\eqref{eq:pertsimplyglobalpoincaretensoreom} reduces to
\begin{equation}\label{eq:hppEOMZeroAngularMomentum}
    {h_{++}}''(z)
    +\frac{\left(-\left(2 z^2+1\right) z^2+2 i \omega  z-3\right)}{z \left(-2 z^2-1\right) \left(z^2-1\right)} {h_{++}}'(z)
    -\frac{4 (\mathcal{J}+1) (\mathcal{J}+2) z+3 i \omega}{z \left(-2 z^2-1\right) \left(z^2-1\right)} {h_{++}}(z) \, .
\end{equation}
Here we have defined the dimensionless radial AdS coordinate $z=r_+/r$.\footnote{This radial $z$-coordinate is not to be confused with the spatial $z$-coordinate used as the third spatial direction in the planar / black brane cases.} 
Compared to the tensor equation of a black brane this equation contains extra terms due to the extra term ``$+1$'' in the blackening factor in Eq.~\eqref{eq:blackeningfactor}. An attempt to solve this equation analytically in a hydrodynamic approximation, at $\mathcal{J}=0$ order by order in $\omega$ following~\cite{Policastro:2002se}, failed. 

Hence, we now consider the large black hole limit of the original equation of motion~\eqref{eq:pertsimplyglobalpoincaretensoreom} for the $h_{++}$-fluctuation around the rotating 5DMPAdS black hole~\eqref{eq:metric5dmpadssig123}. Without any limits, we can numerically solve Eq.~\eqref{eq:pertsimplyglobalpoincaretensoreom} using the standard spectral methods technique discussed in Appendix~\ref{sec:numericalMethods}. The discussion of these numerical results is postponed to the end of this section. Before that, we now gain some analytic insight. The equation of motion~\eqref{eq:pertsimplyglobalpoincaretensoreom} for $h_{++}$, is yet more complicated than Eq.~\eqref{eq:hppEOMZeroAngularMomentum}, which we already failed to solve analytically. Now the coefficients of $h_{++}'(z)$ and $h_{++}(z)$ involve square roots of $a$ and multiple odd as well as even powers of $a$. This seems like a yet more daunting task. However, proceeding fearlessly, we will realize that this equation can be rewritten in a simple and familiar form, namely as the equation of motion obeyed by a massless scalar. 

Now we transform the full equation of motion~\eqref{eq:pertsimplyglobalpoincaretensoreom} so that we arrive in the large black hole case \hyperlink{caseii}{(ii)}, as defined in Sec.~\ref{sec:holographicSetup}. Just like above, we apply the scaling Eq.~\eqref{eq:largeBlackHoleLimit}.
In order to analytically relate our equation of motion to known results from~\cite{Policastro:2002se}, we also perform a transformation to the coordinate $u=z^2$. 
This puts the $h_{++}$-equation into the form of a massless scalar in $AdS_5$-Schwarzschild background known from~\cite{Policastro:2002se}
\begin{equation}\label{eq:pertsimplyglobalpoincaretensoreomtoplanar}
    \begin{aligned}
        \frac{\mathfrak{w}^2-\left(1-u^2\right) \mathfrak{q}^2}{u \left(1-u^2\right)^2} \Phi (u)-\frac{\left(u^2+1\right)}{u(1-u^2)} \Phi '(u)+\Phi ''(u)=0 \, ,
    \end{aligned}
\end{equation}
with a non-trivial definition of angular momentum dependent parameters\footnote{Note that in~\cite{Mamani:2018qzl} a combination of frequency $\nu$ times angular momentum $a$ and the momentum of the perturbation $j$, which we find here, is also observed. There, the authors consider linear momentum. They point out an equivalence between equations of motion in a rotating and a non-rotating background under parameter identifications similar to Eq.~\eqref{eq:identificationsTensorMode}.}
\begin{equation}\label{eq:identificationsTensorMode}
    \begin{aligned}
        \mathfrak{q}^2&:=\frac{L^2 (a\nu - j)^2}{ \left(L^2-a^2\right)}\, ,\\
        \mathfrak{w}^2&:= \frac{\left(L^2 \nu -a j\right)^2}{ \left(L^2-a^2\right)}=\frac{\left(L^2 \bar{\nu}\right)^2}{ \left(L^2-a^2\right)}\, ,\\
        \Phi(u)&:= h_{++}(u) \, ,\\
        f&:=(1-u^2) \, , \\
        \bar{\nu}&:=\nu-a j/L^2 \, ,
    \end{aligned}
\end{equation}
where $\bar{\nu}$ is a convenient frequency to work with 
since it represents the frequency of our modes in a co-rotating frame at the boundary. 
According to Eq.~\eqref{eq:largeBlackHoleLimit}, $\mathfrak{w}$ and $\mathfrak{q}$ each scale like $1/\alpha$ 
so they are small in the large black hole limit we have taken, i.e.~$\mathfrak{w,\,q}\ll 1$. 
Since the analytic solution to this equation in the hydrodynamic limit, $\mathfrak{w,\,q}\ll 1$, is known, we can solve it directly by comparing with~\cite{Policastro:2002se} to obtain the shear correlator at $j=0$ (which implies $\bar\nu=\nu$)
\begin{equation}\label{eq:TppCorrelatorLargeBlackHole}
\langle T^{++} T^{++}\rangle (\nu,\,j=0)
=-i (2\nu) \frac{N^2 \pi^2 T^4}{8} \frac{1}{\sqrt{1-a^2}}
= -i \omega_\text{PSS} \frac{N^2 \pi T^3}{8} \frac{1}{\sqrt{1-a^2}}\, ,
\end{equation}
where we have defined the dimensionless $\nu$ in terms of the frequency, $\omega_\text{PSS} = 2 \pi T \nu L$, 
and $q_\text{PSS} = 2 \pi T j L$, appearing in~\cite{Policastro:2002se}.\footnote{Here, the horizon has been fixed to $r_+=1$ and $L$ has been scaled out through the scaling symmetry in the system.} 
We have checked this correlator numerically and find excellent agreement with our analytic result in Eq.~\eqref{eq:TppCorrelatorLargeBlackHole}. A comparison is shown in Fig.~\ref{fig:ImTppTppOverNu}. 
In the non-rotating limit, Eq.~\eqref{eq:TppCorrelatorLargeBlackHole} reduces to the off-diagonal correlator $\langle T^{x y} T^{x y} \rangle$ in the tensor sector from~\cite{Policastro:2002se}. This shows that at least in the non-rotating limit, the $++$-sector reduces to the $x y$-sector. Both of these sectors decouple from all other sectors. All of these observations give further support for treating $h_{++}$ as the analog of $h_{x y}$, and both may be called~tensor perturbations. We note that it is possible to apply these same steps to the vector equations of motion. This yields a set of equations which we report in the appendix, \eref{eq:pertsimplyglobalpoincarevectoreomtoplanar}, which resembles the vector equations for a Schwarzschild black brane given in~\cite{Policastro:2002se}. However, there are additional terms, which we were not able to get rid of. This hints at the limit \eref{eq:largeBlackHoleLimit} being non trivial. 

Instead of the limit, case \hyperlink{caseii}{(ii)}, we just imposed on the equation of motion, there is another option for taking the large black hole limit, as we already know, leading to the planar limit black brane case \hyperlink{caseii}{(iii)}. In order to reach the latter, one could have taken the large horizon limit according to~\eqref{eq:simpalphalargeTscale} and~\eqref{eq:simpcoordtranstoxyz} on the level of the metric, which we discussed in Sec.~\ref{sec:planarLimitAppendix} referring to it as the {\it planar limit} of the black hole. This has the disadvantage of reducing the geometry to simply being a black brane that has been boosted, as shown in Sec.~\ref{sec:planarLimitAppendix}. Thus, we would lose the properties of rotation. 

Instead, with the goal of preserving the nontrivial aspects of rotating 5DMPAdS perturbations at large temperatures, we have just scaled the linear $h_{++}$ equation of motion in 5DMPAdS, Eq.~\eqref{eq:pertsimplyglobalpoincaretensoreom}, directly, leading to the large black hole case \hyperlink{caseii}{(ii)}. From that, we obtained the analytic solution for the correlator~\eqref{eq:TppCorrelatorLargeBlackHole}. From this, we can define a transport coefficient $\eta_{++}$ in analogy to the shear viscosity $\eta$. We refer to this as the {\it transverse shear viscosity of the spinning black hole} 
\begin{equation}
\label{eq:rigidRotationShearViscosity}
    \eta_{\perp}(a) = \frac{N^2 \pi T_0^3}{8} \frac{1}{\sqrt{1-a^2}}
    = \frac{N^2 \pi T(a)^3}{8} \frac{1}{(1-a^2)^2}\, .
\end{equation}
Divided by the entropy density, this gives the ratio
\begin{equation}
    \frac{\eta_\perp}{s} = \frac{1}{4 \pi}\, ,
\end{equation}
which is the value expected from the general proof for shear modes which satisfy the equation of a massless scalar in Einstein gravity~\cite{Buchel:2003tz, Policastro:2001yc,Kovtun:2003wp,Kovtun:2004de}. This is in agreement with the transverse shear viscosity ratios from the anisotropic Maxwell-dilaton system~\cite{Rebhan:2011vd}, the p-wave superfluid~\cite{Erdmenger:2010xm}, and with the shear viscosity transverse to a strong magnetic field~\cite{Ammon:2020,Critelli:2014kra}. 

In analogy to the longitudinal relation between a shear viscosity and a particular diffusion coefficient, given in Eq.~\eqref{eq:DL}, there should be a diffusion coefficient associated with this transverse shear viscosity $\eta_\perp$. In~\cite{Rebhan:2011vd} the relevant perturbation is given by the sector containing $h_{y z}$ but the momentum has to be chosen perpendicular to the anisotropy, which is the $y$-direction in~\cite{Rebhan:2011vd}. 
However, in our setup, we are not sure how to make such a choice. Hence, this is left for future investigation.

\subsection{Viscosities of planar limit black brane}
\label{sec:viscositiesPlanarLimit}
The transverse shear viscosity of the planar limit black brane can be calculated analytically, while for the longitudinal shear viscosity we obtain a numerical result. Both shear viscosities are identical to the shear viscosities of the simply spinning black hole as a function of the angular momentum parameter $a$.

{First, we analytically compute the transverse shear viscosity of the planar limit black brane.} 
On the planar limit black brane, the helicity-2 perturbation equation~\eqref{eq:pertsimplyglobalpoincaretensoreom} can be solved analytically in the hydrodynamic regime following~\cite{Policastro:2002se}. Setting the momentum to zero for simplicity, we choose the expansion\footnote{Here we note that in Eddington-Finkelstein (EF) coordinates we do not need to factor out a singular contribution as in~\cite{Policastro:2002se}, because the ingoing modes are regular at the horizon in ingoing EF coordinates.} 
\begin{equation}\label{eq:hydroAnsatzhxy}
    h_{x y} = h_{x y}^{(0)} + \omega h_{x y}^{(1)} + \mathcal{O}(\omega^2) \, ,
\end{equation}
Plugging Eq.~\eqref{eq:hydroAnsatzhxy} into the equation of motion for $h_{x y}$ (derived from \eref{eq:pertgenericeom} in the planer limit black brane background metric Eq.~\eqref{eq:metricBoostedBlackBrane}), we obtain two equations of motion, one at zeroth order in $\omega$, the other at linear order in $\omega$. Solving order by order in $\omega$, the resulting solution is valid for all values of the radial coordinate $z$. This solution expanded near the boundary, $z=0$, yields 
\begin{equation}
    h_{x y} = C_0-\frac{i C_0 L^2 \omega  z}{r_+} +\frac{i C_0 L^3 \omega  z^4}{4 r_+ \sqrt{L^2-a^2}}+\mathcal{O}\left(z^5,\omega^2,k\right) \, .
\end{equation}
The vacuum expectation value of the shear tensor component is given by the coefficient of the fourth order term. Varying this with respect to the source, $C_0$, of that same operator gives the correlation function 
\begin{equation}
    \langle T^{x y} T^{x y} \rangle = i \omega
    \frac{N^2 \pi T_0^3}{8}
    \frac{1}{\sqrt{1-a^2}}
    + \mathcal{O}(\omega^2) \, .
\end{equation}
From this, the shear viscosity as a function of the angular momentum parameter, $a$, can be derived with the Kubo formula 
\begin{equation}
\label{eq:planarLimitShearViscosity}
    \eta^{\text{bb}}_{\perp}(a) = \lim\limits_{\omega\to 0}  \frac{\langle T^{x y} T^{x y} \rangle}{i \omega} =
    \frac{N^2 \pi T_0^3}{8}
    \frac{1}{\sqrt{1-a^2}} 
    \, .
\end{equation}
This is the {\it transverse shear viscosity of the planar limit black brane}, i.e.~the shear viscosity of the black brane, Eq.~\eqref{eq:metricBoostedBlackBrane}, in case~\hyperlink{caseii}{(iii)} resulting from the planar limit of the simply spinning black hole. Hence, we have shown analytically, that the transverse viscosities of the rotating planar limit black brane and the simply spinning black hole are identical, i.e.~$\eta^{\text{bb}}_\perp = \eta_\perp$ for all $-1\le a \le 1$. 
It is reassuring that the shear viscosity in this limit agrees with the one from taking the large black hole limit at the level of the fluctuation equation, Eq.~\eqref{eq:rigidRotationShearViscosity}. This also strengthens our hypothesis~\hyperlink{hyp1}{(H.1)} to consider $h_{++}$ as the spinning black hole analog of a black brane tensor perturbation, $h_{x y}$. 
The longitudinal shear viscosity of the planar limit black brane is computed numerically according to the standard Kubo formula
\begin{equation}
    \eta^{\text{bb}}_{||} = \lim\limits_{\omega\to 0}\frac{\langle T_{x z} T_{x z} \rangle (\omega, k=0)}{-i \omega} \, .
\end{equation}
We find that this result coincides (within numerical accuracy) with that of the spinning black hole, i.e. $\eta^{\text{bb}}_{||}=\eta_{||}$ over the whole range of $a$ which we tested.

\section{Discussion}
\label{sec:discussion}
We have analyzed a gravitational setup holographically dual to a spinning fluid, namely spinning black holes. 
We have computed the quasinormal modes~(QNMs) of these large ($r_+\gg L$) five-dimensional Myers-Perry black holes as a function of the angular momentum parameter $-L \le a \le L$ in three distinct cases: choosing large numerical values of $r_+=100$~\hyperlink{casei}{(i)}, performing a strict large horizon limit on the equations of motion~\hyperlink{caseii}{(ii)}, or performing a planar limit on the geometry yielding a black brane~\hyperlink{caseiii}{(iii)}. Here, $|a|=L=1$ 
is the extremal value at which the temperature $T=0$, and the black hole becomes unstable due to superradiance. Our analysis shows that these black holes in the strict large horizon limit \hyperlink{caseii}{(ii)} are stable for all values of $a<L$ against all metric perturbations of the $AdS_5$. 
The spinning black brane~\hypertarget{caseiii}{(iii)} becomes unstable at sufficiently large angular momentum $a\approx 0.75\, L$, see Fig.~\ref{fig:diffusionCoefficient}.

The angular momentum breaks parity symmetry. Therefore, perturbations and observables split into two groups, namely odd and even under parity. This manifests in the QNMs becoming asymmetric about the imaginary frequency axis, i.e.~the symmetry $\nu_{\text{\small QNM}} = -{\nu_{\text{\small QNM}}}^*$ is broken at $a\neq 0$. 
Example plots are shown in Fig.~\ref{fig:hppQNMsofJ} for tensor QNMs, Fig.~\ref{fig:vectqnmsJ} for vector QNMs of the simply spinning black hole as a function of the perturbation's angular momentum parameter $\mathcal{J}$. We have established that this parameter $\mathcal{J}$ and its large black hole analog $j$ defined in \eref{eq:largeBlackHoleLimit} are the rotational analogs of the usual spatial Fourier momentum of perturbations on planar black branes. We have tested this statement using the eikonal limit (large $j$), and comparing the dispersion relations of QNM frequencies $\nu(j)$ in the hydrodynamic limit $\nu, \,j \ll T$ to those known for the black brane. Rotating black holes have some properties in common with the charged Reissner-Nortstr\"om black holes. There is an extremal angular momentum at which the temperature vanishes, just like there is an extremal charge. We observe similarities to the QNMs of RN black branes~\cite{Janiszewski:2015ura} as well. For example, a set of purely imaginary modes enters the system in tensor and vector sector when increasing $a$ from zero towards extremality as seen in Fig.~\ref{fig:hppQNMsofJ} and Fig.~\ref{fig:vectqnmsJ}. This was also the case for tensor and vector QNMs in~\cite{Janiszewski:2015ura}. 

Our focus in this work has been this hydrodynamic regime, i.e.~the lowest QNMs. We have also shown selected results for the non-hydrodynamic quasinormal modes, but those will be discussed in more detail elsewhere~\cite{Garbiso:nonHydroQNMs}. We find that the modes separate into three sectors which are analogous to the non-rotating sectors: tensor sector, momentum diffusion, and sound propagation. 
At nonzero angular momentum, transport coefficients further split into the ones which measure transport along the angular momentum (longitudinal) and the ones measuring transport perpendicular to it (transverse). The following generalized relations are found for the
two shear viscosities, $\eta_\perp$ and $\eta_{||}$, the longitudinal diffusion coefficient, $\mathcal{D}_{||}$, which turns into a damping of a mode now propagating with $v_{||}$, two modified speeds of sound $v_{s,\pm}$ and two sound attenuation coefficients $\Gamma_{\pm}$:
\begin{eqnarray}
 \eta_{\perp}(a) &=& \eta_0 \frac{1}{\sqrt{1-a^2}}\, , \label{eq:etaPerpSum} \\
  \eta_{||}(a) &=& \eta_0 {\sqrt{1-a^2}}\, , \label{eq:etaLSum}\\
  v_{||} & = & a \, , \label{eq:vLSum}\\
  \mathcal{D}_{||} &=& \mathcal{D}_0 (1-a^2)^{3/2} \, ,\\
   v_{s,\pm} & = & v_{s,0} \frac{\sqrt{3} a\pm 1}{1\pm \frac{a}{\sqrt{3}}}\, , \label{eq:vsSum}\\
 \Gamma_{s,\pm} & = & \Gamma_0 \frac{\left(1-a^2\right)^{3/2}}{\left(1\pm \frac{a}{\sqrt{3}}\right)^3} \, ,
\end{eqnarray}
where for the holographic model the quantities at vanishing angular momentum, $a=0$, are: $\eta_0=N^2 \pi T_0^3/8$, $\mathcal{D}_0=1/2$, $v_{s,0}=1/\sqrt{3}$, and $\Gamma_0=1/3$ . 
These quantities are related to each other by generalizations of their $a=0$ Einstein relations
\begin{eqnarray}\label{eq:DLSum}
 \mathcal{D}_{||}(a)
    &=&2 \pi T_0 \frac{\eta_{||}(a)}{\epsilon(a) + P_\perp(a)}\, , \\
    \label{eq:GammaSum}
    \Gamma_\pm(a)&=&\frac{2\eta_{||}(a)}{3(\epsilon(a) + P_\perp(a))} \frac{1}{(1\pm a/\sqrt{3})^3} \, .
\end{eqnarray}
and the bulk viscosity $\zeta(a)=0$ in our model, so we can not make any statement about it. Note that all quantities depend on the angular momentum, with the values given right above (except for the arbitrary normalization constant $(2\pi T_0)$). Of the two pressures in the system, the perpendicular one, $P_\perp$ is relevant for the relations between the parallel shear viscosity and diffusion coefficient, in analogy to~\cite{Rebhan:2011vd}. On top of that, we find that also the two sound attenuations depend on $P_\perp$ and on an additional factor $1/(1\pm a/\sqrt{3})^3$, distinguishing $\Gamma_+$ from $\Gamma_-$. 

{\it Let us stress that we conjecture the hydrodynamic relations Eq.~\eqref{eq:etaPerpSum}, \eqref{eq:etaLSum}, \eqref{eq:vLSum}, \eqref{eq:DLSum}, \eqref{eq:vsSum}, \eqref{eq:GammaSum}, \eqref{eq:DLSum}, and~\eqref{eq:GammaSum} to be valid for any relativistic fluid which has the dimensionless orbital angular momentum $a=\ell/\ell_{\text{critical}}$, where $\ell_{\text{critical}}$ is the orbital angular momentum beyond which superluminal modes would emerge.\footnote{The angular momentum is measured in units of $L$, which is set to one without loss of generality.} We {\it do not} restrict this statement to holographic fluids.}

The ratio of shear viscosity over entropy density varies between zero and $1/(4\pi)$ depending on the direction with respect to the angular momentum in the fluid / black hole. We have computed and displayed the two limiting results, aligned with the angular momentum \eref{eq:specificShearViscosityL} and transverse to the angular momentum \eref{eq:specificShearViscosityT}, see also Fig.~\ref{fig:specificShearViscosities}.  Directions in between these two cases can be parametrized by an angle, which is a generalization of the relations found for a fluid moving linearly~\cite{Kovtun:2019hdm}. 

Our results for the hydrodynamic transport coefficients for momentum diffusion and sound attenuation given in Eqs.~\eqref{eq:vLSum}, \eqref{eq:DLSum}, \eqref{eq:vsSum}, and \eqref{eq:GammaSum}, can be easily understood in terms of reference frames. Consider a fluid at rest being boosted into a rotational motion with angular momentum $a$. 
As a simple example, consider the diffusion mode which becomes a propagating mode with $v_{||}=a\neq 0$. This velocity can be set to zero again if we define the frequency $\bar\nu=\nu-j\, a$ in a co-rotating frame, see \eref{eq:diffusionModeBarNu} and~\cite{Ishii:2018oms}. Generalizing the result from~\cite{Kovtun:2019hdm} to angular motion, the momentum diffusion coefficient $\mathcal{D}_{||}$, the associated velocity $v_{||}$, as well as the sound velocities $v_{s,\pm}$ and attenuations $\Gamma_\pm$ in a rotating fluid can be related to the transport coefficients of a fluid at rest (subscript ``0'') by
\begin{eqnarray} \label{eq:transport0ToTransporta}
    \nu &=& a\, j - i \mathcal{D}_0 \sqrt{1-a^2} (2 \pi T)j^2(1-a^2)+\mathcal{O}(j^3)\, , \quad \text{(diffusion)}\\ 
    \nu &=& \frac{a\pm v_{s,0}}{1\pm a v_{s,0}} j - \frac{i}{2} {\Gamma_0} \frac{(1-a^2)^{3/2}}{(1\pm a \, v_{s,0})^3} (2\pi T) j^2 + \mathcal{O}(k^3) \, ,  \quad \text{(sound)}
    \label{eq:transport0ToTransportaSound}
\end{eqnarray} 
where the angular momentum of the perturbation with magnitude $j$ is aligned with the angular momentum of the fluid with magnitude $a$. 
These relations are motivated and discussed below \eref{eq:movingFluidDiffusion} and \eref{eq:movingFluidSound}. We have demonstrated validity of these relations by explicit calculation in our holographic model. 
Equation~\eqref{eq:transport0ToTransporta} is further verified by comparison to the hydrodynamic sound and shear diffusion modes found in $AdS_4$ rotating black strings~\cite{Mamani:2018qzl}. Recall, that in the latter work, the authors consider a spacetime which has the topology of a cylinder but is locally a boosted black brane. There, the authors study hydrodynamic modes propagating with linear momentum along or perpendicular to the angular momentum of the fluid. Setting the perpendicular momentum of their perturbations to zero, keeping only the longitudinal one, we confirm exact agreement with our terms linear in $j$ in~\eqref{eq:transport0ToTransporta} in both, diffusion and sound modes. The term quadratic in $j$ in the diffusion mode and sound mode also agree. Of course~$\mathcal{D}_0$ and $v_{s,0}$ in~\cite{Mamani:2018qzl} take the values for a (2+1)-dimensional non-rotating fluid, whereas in our case those are the (3+1)-dimensional values. 

Taking a step back, once again we recognize that the splitting between longitudinal and transverse transport effects is a universal feature of all anisotropic systems described by hydrodynamics. 
The exact same splitting we observe in this work has been derived for intrinsically anisotropic plasma in~\cite{Rebhan:2011vd}. 
This type of splitting has further been shown in the case in which a weak external magnetic field breaks isotropy~\cite{Ammon:2017ded,Abbasi:2015saa,Abbasi:2016rds,Abbasi:2017tea,Kalaydzhyan:2016dyr}. 
In the same way, transport coefficients in the presence of strong dynamical magnetic fields have been revealed in holographic studies~\cite{Grozdanov:2017kyl} based on the formulation of magnetohydrodynamics in~\cite{Grozdanov:2016tdf}. Specifically, in~\cite{Grozdanov:2017kyl} the longitudinal shear viscosity and longitudinal bulk viscosity have been determined alongside the longitudinal and transverse charge resistivities. As a feature in common with our study, at large anisotropies, regardless whether caused by magnetic field or rotation, the longitudinal shear viscosity asymptotes to zero while the transverse shear vicosity is unaffected. In agreement with this behavior, our longitudinal diffusion coefficient, $\mathcal{D}_{||}$, also vanishes at large anisotropy, see Fig.~\ref{fig:diffusionCoefficient}. This exact behavior is observed in strong static magnetic fields~\cite{Finazzo:2016mhm}, showing that the dynamics of the gauge fields is irrelevant for this observation. In contrast to that, curiously {\it both} charge resistivities appear to asymptote to zero in~\cite{Grozdanov:2017kyl}, apparently not distinguishing between longitudinal and transverse directions at strong anisotropy. It would be interesting to understand this distinct behavior.  
In summary we recall that the longitudinal shear viscosity is an excellent example showing that transport is severely different in the presence of anisotropies. The seemingly universal isotropic value of $\eta/s=1/(4\pi)$ is not realized for $\eta_{||}/s\le 1/(4\pi)$. As discussed based on technical details in Sec.~\ref{sec:longituidnalShear}, the reason for this deviation is that the anisotropy breaks rotational invariance in the plane in which the shear is applied. 
As stated in the introduction, these effects of anisotropies will considerably affect numerical implementations of hydrodynamics in the endeavor of describing the quark gluon plasma. More generally the effects we have demonstrated here have to be considered for any anisotropic system, e.g.~to describe transport in anisotropic condensed matter materials. 
 
An obvious continuation of our work is to study the higher QNMs in more detail~\cite{Garbiso:nonHydroQNMs}. Another future task is to understand the shifted asymptotic behavior of QNMs in the large momentum (eikonal) limit visualized in Fig.~\ref{fig:hpplargej}. It would be useful to understand if and how the Wigner-D representations labeled by half-integer valued $\mathcal{J}$ in the large black hole limit~\eref{eq:largeBlackHoleLimit} can be written in a closed form as a function of the continuous angular momentum $j$. This can confirm hypothesis~\hyperlink{hyp1}{(H.I)}. This may help with a rigorous derivation of \eref{eq:transport0ToTransporta} along~\cite{Kovtun:2019hdm}. Hypothesis~\hyperlink{hyp2}{(H.II)} should be confirmed by an explicit derivation of Kubo formulae for all transport coefficients from the hydrodynamic constitutive equations for a rotating fluid in analogy to~\cite{Hernandez:2017mch,Kovtun:2018dvd,Niemi:2011ix,Ammon:2020}. The results can be cross-checked with a holographic calculation in the setup we presented here. 
The algebraic methods applied in~\cite{Lysov:2017cmc} may prove useful for this task. 
Finally, it would be a long term goal to rigorously couple hydrodynamics of rotating fluid to spin degrees of freedom in order to understand the hyperon polarization observed in heavy ion collisions~\cite{STAR:2017ckg}, and potentially other spin and spin transport effects in the QGP. Fluid/gravity approaches seem most promising for this task, along~\cite{Gallegos:2020otk}, or extending~\cite{Bhattacharyya:2007vs}. 
An interesting complementary view of rotating black holes would emerge if one considered the system at hand in the extremal case $a=1$~\cite{Edalati:2010hk,Moitra:2020dal}. 
Exact results may be obtained from 
a recently proposed method which applies an exact version of Bohr-Sommerfeld quantization conditions, yielding exact quasinormal mode frequencies~\cite{Aminov:2020yma}. Exact equations for diffusion coefficients and shear viscosities as functions of the background metric components using the membrane paradigm could be investigated~\cite{Kovtun:2003wp}, similar to~\cite{Rebhan:2011vd}, or using analytic considerations for the lowest QNMs~\cite{Starinets:2008fb}. 
Although in this work we were not able to identify an appropriate perturbation, a transverse diffusion coefficient $\mathcal{D}_\perp$, should be present in our system, associated with the transverse shear viscosity. Rotating non-relativistic fluids would be interesting for condensed matter applications, e.g.~in the framework of a non-relativistic limit of the relativistic hydrodynamic formulation~\cite{Kaminski:2013gca,Jensen:2014wha}, or with the help of Ho\v rava gravity~\cite{Garbiso:2019uqb,Davison:2016auk}. Introducing time-dependence into charged magnetic black branes far from equilibrium would provide a holographic model of thermalization and transport including the major features of an evolving quark gluon plasma~\cite{Janik:2005zt,Chesler:2008hg,Chesler:2010bi,Cartwright:2019opv,Cartwright:2020qov,Wondrak:2020tzt,Wondrak:2017kgp,Endrodi:2018ikq}.
\section*{Acknowledgements} 
We thank your group members Casey Cartwright, Jana Ingram, Marco Knipfer, and Roshan Koirala for many discussions. We thank Pavel Kovtun, Andrei Starinets, and Allen Stern for valuable comments. 
MG would like to thank Takaaki Ishii and Keiju Murata for insightful discussions and for pointing out several problems with calculating thermodynamic quantities. 
MG would also like to thank Gi-Chol Cho, Shin Nakamura, and Tatsuma Nishioka for 
hospitality and discussions. 
MG acknowledges financial support from the University of Alabama Graduate School, facilitating his research visit to Japan where part of this work was conducted. 
This  work  was supported,  in  part,  by  the  U.S.~Department of Energy grant DE-SC-0012447.

\appendix

\section{Numerical methods} 
\label{sec:numericalMethods}
\subsection{Quasinormal modes}
We use a pseudospectral grid to discretize our equations of motion (see \cite{Piotrowska:2017pwe} for a deeper dive) (see \cite{Jansen:2017oag} for a similar numerical routine) where the perturbations are ingoing at the horizon and obey a Dirichlet boundary condition at the conformal boundary. Alternatively, one may start with the metric in Ingoing Eddington Finkelstein coordinates where one only has to make sure that the perturbations obey the Dirichlet boundary condition at the conformal boundary. For purposes of illustrating the numerical method used, we will assume the perturbation fields, $\varphi_i(z)$, are regular functions of an inverted, scaled radial coordinate,$z=r_+/r$, that vanish at the conformal boundary ($z=0$). In general $\varphi_i(z)$ are solutions to the set of second order ordinary differential equations, $\left( A_{ij}\partial_z^2+B_{ij}\partial_z+C_{ij} \right)\varphi_j=0$ where matrices are function of $z$ valued. We may discretize this equations by writing $\varphi_i$  as $(\vec{\varphi}_i)_k=\varphi(z_k)_i$ where $\vec{\varphi}_i$ is a column vector where the components are $\vec{\varphi}_i$ evaluated on the kth Chebyshev-Gauss-Lobatto gridpoint. Similarly we write the components of $A$, $D$, and $C$ as diagonal matrices where the k-k th component of each of these matrices is evaluated on the kth gridpoint. The derivative are written as matrices given by \cite{Piotrowska:2017pwe}. One then ends up with linear algebra problem, $\mathcal{A}\mathfrak{v}=0$. We then write the problem with the dependence on $\omega$ explicitly express, $\left( \sum_{n=0}^{N} \omega^n \mathcal{A} \right) \mathfrak{v} = 0$. Introducing the appropriate number of auxiliary fields, $\mathfrak{y}_i=\omega \mathfrak{y}_{i+1}$ and $\mathfrak{v}=\omega \mathfrak{y}_{1}$, where one has a block matrix matrix equation $\mathbb{A} \mathcal{V}=\omega \mathbb{B} \mathcal{V}$. This is a general eigenvalue problem where the eigenvalues are the quasinormal frequencies. 

\subsection{Holographic correlation functions}
A proper discussion of holographic correlation functions can be found in~\cite{Skenderis:2008dg,Skenderis:2008dh} and in particular for coupled systems in~\cite{Amado:2009ts,Kaminski:2009dh}. Here we summarize the concepts and steps in the calculation. With the choice of appropriate divergent and non-divergent counter-terms, a holographic correlation function can be written as
\label{sec:correlators}
\begin{equation}
    \langle \mathcal{O}_1 \mathcal{O}_2\rangle = \frac{\langle \mathcal{O}_1 \rangle}{\phi_2^{(0)}} \, ,
\end{equation}
where the vacuum expectation value~(vev) $\langle \mathcal{O}_1 \rangle$ and source $\phi_2^{(0)}$ can be extracted from the near boundary expansion of the gravitational fields $\phi_1$ and $\phi_2$ which are dual to the operators $\mathcal{O}_2$ and $\mathcal{O}_2$, respectively. For example, for components of operators with operator dimension 4, such as the energy momentum tensor, the near boundary expansions are given by this\footnote{When the two solutions (normalizable and non-normalizable) for one field differ by an integer, logarithmic terms appear at the order in $z$ which equals the operator dimension of the dual operator. Here we do not display them for the sake of simplicity. However, we take logarithmic terms into account in all our computations.}
\begin{eqnarray}
 \phi_1&=& \phi_1^{(0)} + \dots + z^4 \langle \mathcal{O}_1 \rangle  + \dots \, , \\
 \phi_2&=& \phi_2^{(0)} + \dots + z^4 \langle \mathcal{O}_2 \rangle + \dots  \, .
\end{eqnarray}
Note that in general the vevs depend on both sources if the operators mix, or equivalently if the dual fields are coupled, i.e. we have 
$\langle \mathcal{O}_1\rangle 
= \langle \mathcal{O}_1\rangle (\phi_1^{(0)},\phi_2^{(0)})$ and 
$\langle \mathcal{O}_2\rangle 
= \langle \mathcal{O}_2 \rangle (\phi_1^{(0)},\phi_2^{(0)})$. 
So for the energy momentum tensor components this would read 
\begin{eqnarray}
 h_{+3}&=& h_{+3}^{(0)} + \dots + z^4 \langle T_{+3} \rangle (h_{+3}^{(0)} ,h_{+t}^{(0)} )  + \dots \, , \\
 h_{+t}&=& h_{+t}^{(0)} + \dots + z^4 \langle T_{+t} \rangle (h_{+3}^{(0)} ,h_{t3}^{(0)} )  + \dots  \, .
\end{eqnarray}
Now we intend to only source one operator on the boundary in order to extract its correlation function with itselve or with another operator. This is achieved by choice of boundary conditions on the sources. For example, we compute $\langle T_{+t} T_{+t} \rangle = \langle T_{+t}\rangle/h_{+t}^{(0)}$ from the solution to the coupled fluctuation equations with boundary conditions $\{h_{+3}^{(0)}=0,h_{+t}^{(0)}=1\}$. 

\section{Perturbation equations} 
\label{sec:perturbationEqs}
In this paper we are concerned with the first order hydrodynamic and non-hydrodynamic response behavior of the rotating black hole and of the holographically dual fluid. We analyze the three channels in 5D: shear (tensor), momentum diffusion (vector), and sound propagation (scalar). 
For the sake of calculating QNFs and near horizon behavior, we will use ingoing Eddington-Finkelstein coordinates where
\begin{equation}
    \begin{aligned}
        & dt = dv +\frac{L^2 r^2 \sqrt{a^2 \left(L^2+r_+^2\right)-L^2 r_+^2} \sqrt{a^2 \left(L^2+r_+^2\right) \left(r^4-r_+^4\right)-L^2 r^4 r_+^2}}{a^2 \left(L^2+r^2\right) \left(L^2+r_+^2\right) \left(r^4-r_+^4\right)-L^2 r^2 r_+^2 (r-r_+) (r+r_+) \left(L^2+r^2+r_+^2\right)} dr \\
        & \sigma^3 = \tilde{\sigma}^3+\frac{2 a L^2 r^2 r_+^4 \left(L^2+r_+^2\right) \sqrt{a^2 \left(L^2+r_+^2\right)-L^2 r_+^2}}{\mathcal{F}(r)} dr \, , \\
        & \mathcal{F}(r) = \left(r^2-r_+^2\right) \left(a^2 \left(L^2+r^2\right) \left(L^2+r_+^2\right) \left(r^2+r_+^2\right)-L^2 r^2 r_+^2 \left(L^2+r^2+r_+^2\right)\right)\times\\
        &\sqrt{a^2 \left(L^2+r_+^2\right) \left(r^4-r_+^4\right)-L^2 r^4 r_+^2} \, .
    \end{aligned}
    \label{eq:framesimplyedfinkelstein}
\end{equation} 
\paragraph{Momentum Diffusion Channel}
Plugging in (\ref{eq:pertsimplyglobalvector}) into (\ref{eq:pertgenericeom}) we have the following equations of motion for ($h_{+t}$,$h_{+3}$,$h_{++}$):
\pagestyle{empty}
\begin{equation}\label{eq:pertsimplyglobalpoincarevectoreom}
    \begin{aligned}
        0 = & h_{t+}''(r) + \frac{L^2 \left(2 a^2 \mu -10 \mu  r^2+5 r^4\right)+2 a^2 \mu  r^2+5 r^6}{L^2 \left(2 a^2 \mu  r-2 \mu  r^3+r^5\right)+2 a^2 \mu  r^3+r^7}h_{t+}'(r) + \\
        &\frac{8 a \mu  \left(L^2+2 r^2\right)}{L^2 \left(2 a^2 \mu  r-2 \mu  r^3+r^5\right)+2 a^2 \mu  r^3+r^7} h_{3+}'(r)+\\
        &\frac{L^2 h_{t+}(r)}{\left(L^2 \left(2 a^2 \mu  r-2 \mu  r^3+r^5\right)+2 a^2 \mu  r^3+r^7\right)^2} (-4 L^2 (4 a^4 \mu ^2-2 a^2 \mu ^2 r^2 (a \omega -2 \mathcal{J})\\
        &+\mathcal{J} \mu  r^6 (a \omega -2 \mathcal{J}-4)+\mathcal{J} (\mathcal{J}+2) r^8)-16 a^4 \mu ^2 r^2-4 \mathcal{J} (\mathcal{J}+2) r^{10})-\\
        &\frac{2 i \sqrt{2} \sqrt{\mathcal{J}} L^2 r^2 \omega }{L^2 \left(2 a^2 \mu -2 \mu  r^2+r^4\right)+2 a^2 \mu  r^2+r^6} h_{++}(r)-\\
        &\frac{4 L^2 h_{3+}(r)}{\left(L^2 \left(2 a^2 \mu  r-2 \mu  r^3+r^5\right)+2 a^2 \mu  r^3+r^7\right)^2} (-8 a^3 \mu ^2 \left(L^2+r^2\right)-2 a^2 \mu  r^2 \omega  \left(L^2 \left(2 \mu +r^2\right)+r^4\right)+\\
        &a \mu  r^2 \left(L^2 r^4 \omega ^2-4 (\mathcal{J}+2) \left(L^2 \left(r^2-2 \mu \right)+r^4\right)\right)+\mathcal{J} r^6 \omega  \left(L^2 \left(r^2-2 \mu \right)+r^4\right))\text{,}\\
        0 = & h_{++}''(r) + \frac{L^2 \left(3 r^4-2 \mu  \left(a^2+r^2\right)\right)+2 a^2 \mu  r^2+5 r^6}{L^2 \left(2 a^2 \mu  r-2 \mu  r^3+r^5\right)+2 a^2 \mu  r^3+r^7} h_{++}'(r) -\\
        &\frac{2 i \sqrt{2} \sqrt{\mathcal{J}} L^4 r^4 \left(2 a \mu  (a \omega -2 \mathcal{J}-2)+r^4 \omega \right)}{\left(L^2 \left(2 a^2 \mu -2 \mu  r^2+r^4\right)+2 a^2 \mu  r^2+r^6\right)^2} h_{t+}(r)+\\
        & \frac{h_{++}(r)}{\left(L^2 \left(2 a^2 \mu -2 \mu  r^2+r^4\right)+2 a^2 \mu  r^2+r^6\right)^2}(-4 (\mathcal{J}+1) L^2 r^2 (\mathcal{J} r^4 \left(L^2+r^2\right)-\\
        &2 \mu  \left(a^2 \left(L^2+r^2\right)+\mathcal{J} L^2 r^2\right))+L^4 r^4 \omega ^2 \left(2 a^2 \mu +r^4\right)-8 a (\mathcal{J}+1) \mu  L^4 r^4 \omega ) +\\
        & \frac{8 i \sqrt{2} \sqrt{\mathcal{J}} L^2 r^4 \left(a \mu  L^2 \omega +(\mathcal{J}+1) \left(L^2 \left(r^2-2 \mu \right)+r^4\right)\right)}{\left(L^2 \left(2 a^2 \mu -2 \mu  r^2+r^4\right)+2 a^2 \mu  r^2+r^6\right)^2} h_{3+} \text{,}\\
        0 = & h_{3+}''(r) -\frac{4 a \mu  L^2 r}{L^2 \left(2 a^2 \mu -2 \mu  r^2+r^4\right)+2 a^2 \mu  r^2+r^6} h_{t+}'(r) +\\
        & \frac{L^2 \left(6 a^2 \mu -2 \mu  r^2+3 r^4\right)+5 r^2 \left(2 a^2 \mu +r^4\right)}{L^2 \left(2 a^2 \mu  r-2 \mu  r^3+r^5\right)+2 a^2 \mu  r^3+r^7} h_{3+}'(r) -\\
        &\frac{L^4 \left(\mathcal{J} r^4-2 a^2 \mu \right) \left(2 a \mu  (a \omega -2 \mathcal{J}-2)+r^4 \omega \right)}{\left(L^2 \left(2 a^2 \mu -2 \mu  r^2+r^4\right)+2 a^2 \mu  r^2+r^6\right)^2} h_{t+}(r)-\\
        & \frac{2 i \sqrt{2} \sqrt{\mathcal{J}} (\mathcal{J}+1) L^2 r^2}{L^2 \left(2 a^2 \mu -2 \mu  r^2+r^4\right)+2 a^2 \mu  r^2+r^6} h_{++}(r)+\\
        & \frac{h_{+3}(r)}{\left(L^2 \left(2 a^2 \mu  r-2 \mu  r^3+r^5\right)+2 a^2 \mu  r^3+r^7\right)^2} (L^4 (-32 a^4 \mu ^2-8 a^2 \mu ^2 r^2 (a \omega -2 \mathcal{J}-6)-\\
        &16 a^2 (\mathcal{J}+2) \mu  r^4+2 \mu  r^6 (-2 \mathcal{J} (a \omega -4)+a \omega  (a \omega -4)+8)\\
        &-8 (\mathcal{J}+1) r^8+r^{10} \omega ^2)-8 L^2 r^2 \left(2 a^2 \mu +r^4\right) \left(2 a^2 \mu +(\mathcal{J}+1) r^4\right))  \text{,}\\
        0 = & h_{3+}'(r)-\frac{i \sqrt{\mathcal{J}} \left(L^2 \left(2 a^2 \mu -2 \mu  r^2+r^4\right)+2 a^2 \mu  r^2+r^6\right)}{\sqrt{2} \left(L^2 \left(-2 a^2 \mu +\mu  r^2 (a \omega -2 \mathcal{J})+\mathcal{J} r^4\right)-2 a^2 \mu  r^2+\mathcal{J} r^6\right)} h_{++}'(r)-\\
        &\frac{L^2 r^2 \left(2 a \mu  (a \omega -2 \mathcal{J}-2)+r^4 \omega \right)}{4 \left(L^2 \left(-2 a^2 \mu +\mu  r^2 (a \omega -2 \mathcal{J})+\mathcal{J} r^4\right)-2 a^2 \mu  r^2+\mathcal{J} r^6\right)} h_{t+}'(r) + \\
        &\frac{4 a \mu  L^2 r}{L^2 \left(-2 a^2 \mu +\mu  r^2 (a \omega -2 \mathcal{J})+\mathcal{J} r^4\right)-2 a^2 \mu  r^2+\mathcal{J} r^6} h_{t+}(r)-\\
        &\frac{8 a^2 \mu  \left(L^2+r^2\right)}{L^2 \left(-2 a^2 \mu  r+\mu  r^3 (a \omega -2 \mathcal{J})+\mathcal{J} r^5\right)-2 a^2 \mu  r^3+\mathcal{J} r^7} h_{3+}(r) \, .
    \end{aligned}
\end{equation} 
We have chosen the standard holographic radial gauge here so we only have three dynamical equations and one constraint. For the special value of $\mathcal{J}=0$ the ``$++$'' component of the Einstein equations decouples from this sector. This decoupling also occurs in the large black hole limit, case~\hyperlink{caseii}{(ii)}, for all values of $j$.

\paragraph{Sound channel} We do not report the coupled scalar perturbation equations in this work due to excessive length. We are happy to share these equations with the interested reader upon request.

\paragraph{Large black hole limit of vector perturbation equations}
We perform the same limiting procedure, \eref{eq:largeBlackHoleLimit}, to the momentum diffusion channel as was applied to the tensor equation of motion. We find the following
\begin{equation}\label{eq:pertsimplyglobalpoincarevectoreomtoplanar}
    \begin{aligned}
    H_t'+\frac{2 (a \mathfrak{w}+f L \mathfrak{q})}{L (a f \mathfrak{q}+L \mathfrak{w})} H_3' & = 0\text{,}\\
    H_t''-\frac{f L^2-a^2 (u^2+1)}{f u \left(L^2-a^2\right)}H_t' + \frac{4 a u}{f \left(L^2-a^2\right)}H_3' & - \\ \frac{2}{L} \frac{\left(a \mathfrak{w}+f L \mathfrak{q}\right)}{f^2 u \left(L^2-a^2\right)} \left(\frac{L}{2} (a \mathfrak{w}+L \mathfrak{q}) H_t+(a \mathfrak{q}+L \mathfrak{w}) H_3\right) & = 0\text{,}\\
    H_3''-\frac{a L^2 u}{f \left(L^2-a^2 \right)} H_t' - \frac{L^2 \left(u^2+1\right)-a^2 f}{f u \left(L^2-a^2\right)} H_3' & + \\ \frac{\left(a \mathfrak{q} f+L \mathfrak{w}\right)}{f^2 u \left(L^2-a^2\right)} \left(\frac{L}{2} (a \mathfrak{w}+L \mathfrak{q}) H_t+(a \mathfrak{q}+L \mathfrak{w}) H_3\right) & = 0
    \end{aligned}
\end{equation}
where 
\begin{equation}
    \begin{aligned}
        H_3(u) & := - h_{+3}(u)\\
        H_t(u) & := h_{+t}(u) \, .
    \end{aligned}
\end{equation}
In this limit, the equation of motion for $h_{++}$ in this vector sector is the the same as in the tensor sector and decouples from the other vector fields, so we will not report it here. Unlike \eref{eq:pertsimplyglobalpoincaretensoreomtoplanar}, \eref{eq:pertsimplyglobalpoincarevectoreomtoplanar} can not be taken to the form of the nonrotating case due to non-trivial terms that depend on u. Nevertheless, \eref{eq:pertsimplyglobalpoincarevectoreomtoplanar} does agree with \cite{Policastro:2002se} when $a=0$.

\section{Planar limit black brane} 
\label{sec:planarLimitAppendix}
\begin{figure}
    \centering
    \includegraphics[width=0.8\textwidth]{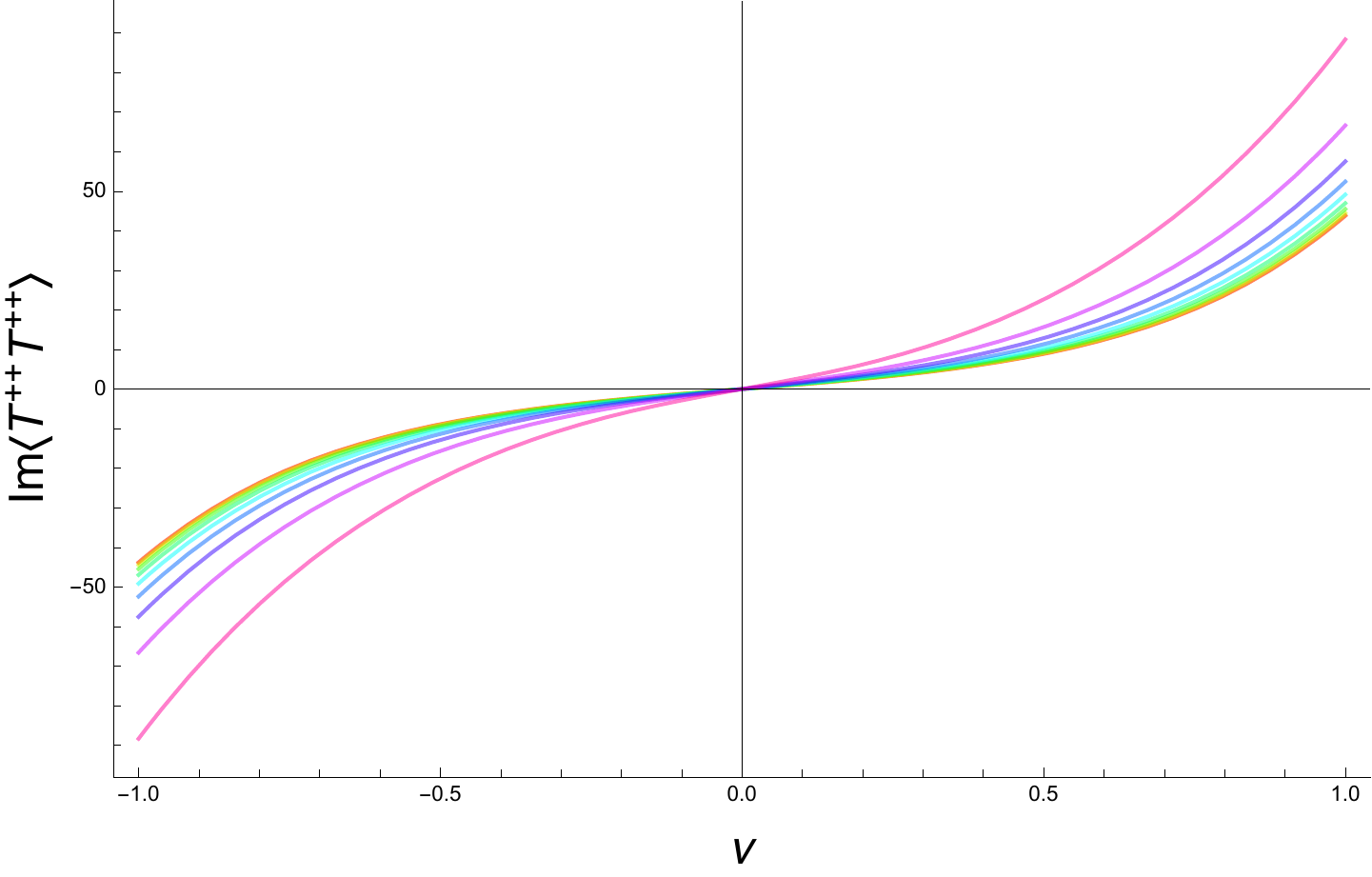}
    \caption{\label{fig:ImTppTppOverNu}
    Numerical result for $\text{Im}\langle T^{++} T^{++}\rangle$ displayed as a function of $\nu$, where the color encodes the values of $a$ in the range $0<a<1$. Red is the lower bound while purple is the higher bound.
}
\end{figure}

\begin{figure}
    \centering
    \includegraphics[width=0.8\textwidth]{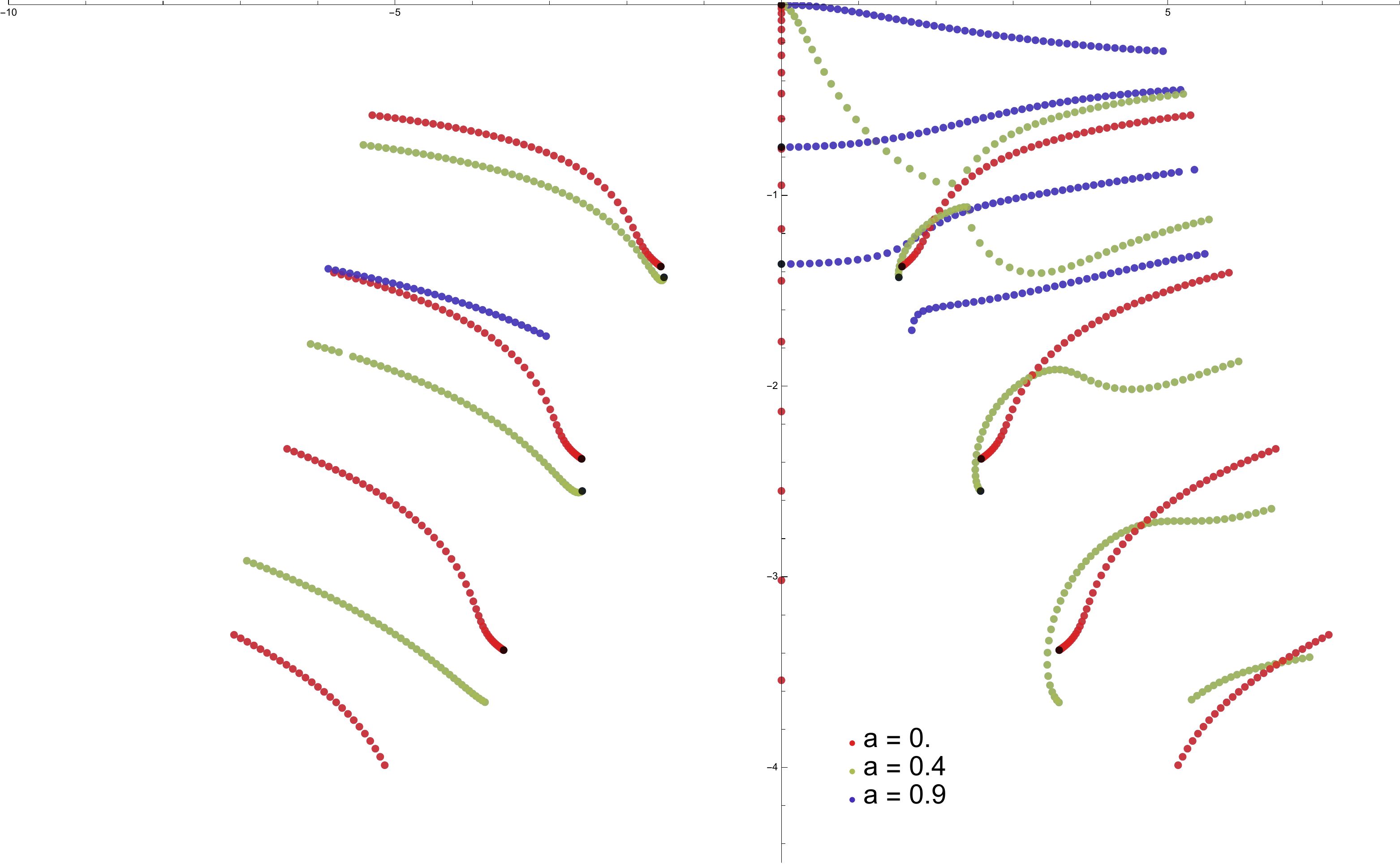}
    \caption{The vector QNM frequencies shown in Fig.~\ref{fig:qnfsh3+t+} but displayed in the complex frequency plane. Shown are several values of $j$ where the black point on each curve indicates $j=0$.}
    \label{fig:qnfsh3+t+incompplane}
\end{figure}

Since there is $\text{ISO}(2)$ symmetry in the ($x-y$) plane and translation symmetry (before the boost \eref{eq:boostTrafo}) in the $\tau$ and  $z$ directions we can construct the following perturbations of \eref{eq:metricBoostedBlackBrane}) as \eref{eq:pertgeneric}:
\begin{equation}\label{eq:pertplanargeneric}
    \begin{aligned}
        h_{\text{brane}}= \int d\omega d{k} ~ r^2 h_{\mu \nu}(\omega,k;r) e^{i(-\omega \tau + k z)} \, ,
    \end{aligned}
\end{equation}
where we will suppress the $\omega$ and $k$ dependence. Furthermore, we can take the radial gauge where $h_{\mu r} = 0$. Since $\text{O}(2) \subset \text{ISO}(2)$ the perturbation's components decouple in (\ref{eq:pertgenericeom}) according to how they transform under $\text{O}(2)$. Here we collect the resulting QNMs in example figures~\ref{fig:ImTppTppOverNu}, \ref{fig:qnfsh++}, and~\ref{fig:qnfsh3+t+incompplane}. 
\begin{figure}
    \centering
    \includegraphics[width=0.8\textwidth]{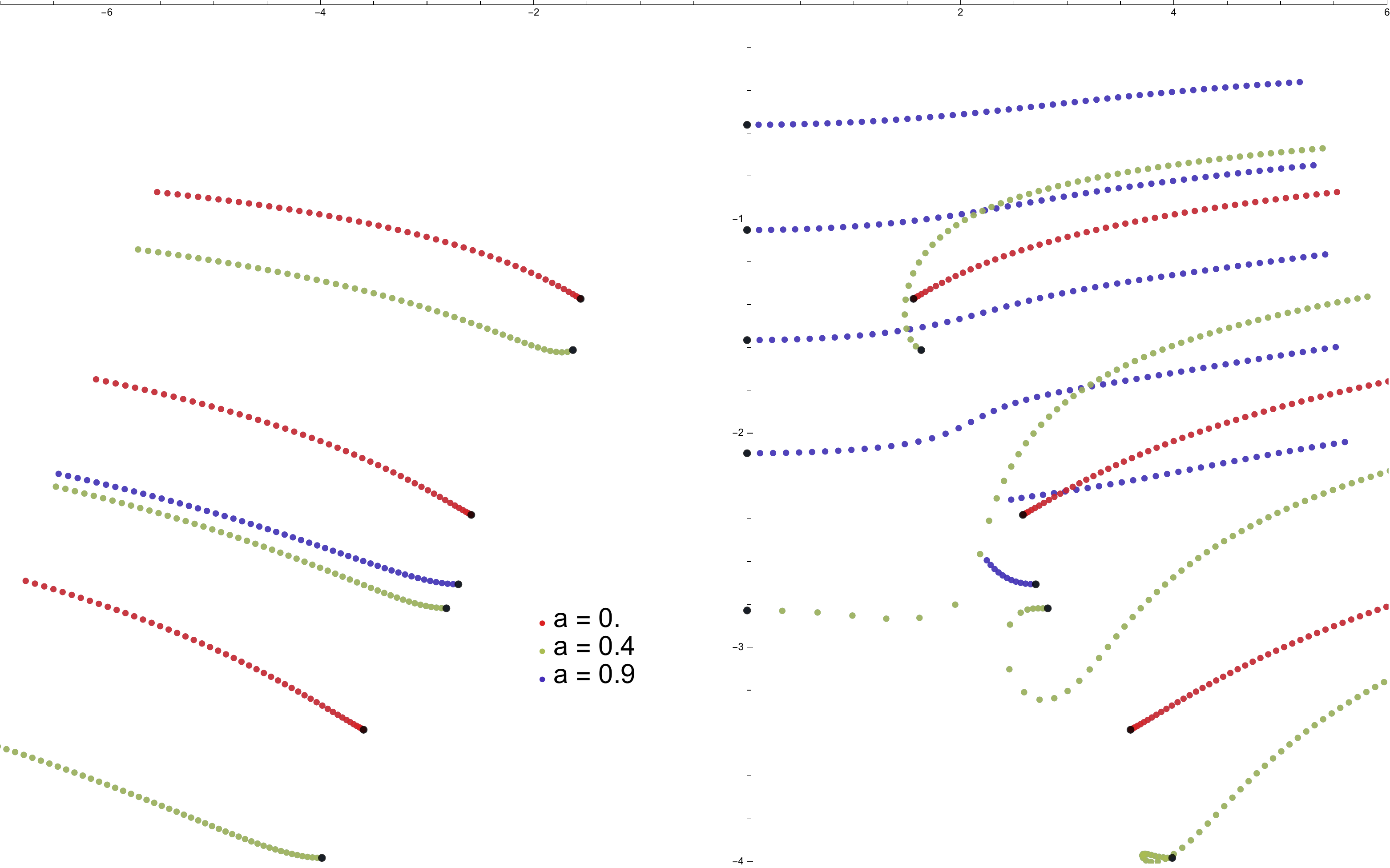}
    \caption{Tensor QNM frequencies as in Fig.~\ref{fig:qnfsh++} but displayed in the complex frequency plane. Shown are several values of $j$ where the black point on each curve indicates $j=0$.}
    \label{fig:qnfsh++incompplane}
\end{figure}

\bibliographystyle{JHEP}
\bibliography{bib.bib}

\end{document}